\documentclass[sensors,article,accept,moreauthors,pdftex,12pt,a4paper]{mdpi}

\lastpage{x}
\doinum{10.3390/------}
\pubvolume{15}
\pubyear{2015}
\externaleditor{Academic Editor: Leonhard M. Reindl}
\history{Received: 25 February 2015 / Accepted: 16 April 2015 / Published: xx}

\usepackage[english]{babel}
\usepackage[latin1]{inputenc}
\usepackage{graphicx}
\usepackage{algorithm}
\usepackage{algorithmic}
\usepackage{amsfonts}
\usepackage{amssymb}
\usepackage{mathtools}
\usepackage{amsmath}
\usepackage{bm}
\usepackage{soul}
\usepackage{booktabs}
\usepackage{enumitem}
\usepackage{upgreek}

\DeclarePairedDelimiter\floor{\lfloor}{\rfloor}

\newcommand{\FUNCNAME}[1]{\textsc{#1}}

\Title{Online Learning Algorithm for Time Series Forecasting Suitable
    for Low Cost Wireless Sensor Networks Nodes}
	
\Author{Juan Pardo *, Francisco Zamora-Mart\'inez and Paloma Botella-Rocamora}

\address[1]{%
ESAI---Embedded Systems and Artificial Intelligence Group,
Escuela Superior de Ense\~nanzas T\'ecnicas, Universidad CEU Cardenal Herrera, C/San~Bartolom\'e, 46115 Alfara del Patriarca, Valencia, Spain; E-Mails:~francisco.zamora@uch.ceu.es~(F.Z.M.); pbotella@uch.ceu.es (P.B.R.)\vspace{-12pt}}

\corres{E-Mail: juan.pardo@uch.ceu.es; \linebreak Tel.: +34-961-369-000 (ext 2363).}

\abstract{Time series forecasting is an important predictive methodology which
  can be applied to a wide range of problems.~Particularly, forecasting the
  indoor temperature permits an improved utilization of the HVAC (Heating,
  Ventilating and Air Conditioning) systems in a home and thus a better energy
  efficiency.~With such purpose the paper describes how to implement an
  Artificial Neural Network (ANN) algorithm in a low cost system-on-chip to
  develop an autonomous intelligent wireless sensor network.~The present paper
  uses a Wireless Sensor Networks (WSN) to monitor and forecast the indoor
  temperature in a smart home, based on low resources and cost microcontroller
  technology as the 8051MCU.~An on-line learning approach, based on
  Back-Propagation (BP) algorithm for ANNs, has been developed for real-time time
  series learning.~It performs the model training with every new data that
  arrive to the system, without saving enormous quantities of data to create a
  historical database as usual, \emph{i.e.}, without previous knowledge.~Consequently to
  validate the approach a simulation study through a Bayesian baseline model
  have been tested in order to compare with a database of a real application aiming to see the
  performance and accuracy.~The core of the paper is a new algorithm, based on the BP one, which has
  been described in detail, and the challenge was how to implement a
  computational demanding algorithm in a simple architecture with very few hardware resources.}


\begin{document}

\maketitle

\section{Introduction}

Wireless Sensor Networks (WSNs) have been widely considered as one of the most
promising present and future technologies.~In fact, the latest advances in wireless communication
technologies have made it possible to develop tiny, cheap and smart sensors
embedded in a small physical area, with wireless network capabilities, that
provides huge opportunities for a vast variety of applications.~Some common examples can
be found, such as industrial monitoring processes, machine health monitoring, physical and
environmental conditions monitoring, \emph{etc.}~\cite{zheng2009wireless}.~However, one of its most promising
applications is on smart homes and ambient intelligence, which makes it feasible to provide
scalable intelligent networks of sensors/actuators according to new home
technologies appear on the market.~WSNs can be used to provide more convenient
and intelligent living environments for human beings and can be embedded into a
house to develop an autonomous home network.~The present paper uses a WSN to
monitor and forecast the indoor temperature in a smart home, based on low
resources and low cost microcontroller~technology.

Several studies say that in the European Union about 40\% of total primary
energy demand corresponds to buildings'  consumption~\cite{Ferreira}.~At home, more
than a half of such consumption is produced by HVAC (Heating, Ventilating and
Air Conditioning) systems~\cite{Alvarez}.~The indoor temperature is the most crucial
variable that determines the utilization of such systems and thus has a major
effect on the overall energy expenditure.~For that reason, it is still necessary
to develop new intelligent systems at home to manage the demand of energy
efficiently, considering a plausible balance between consumption and
comfort.~To develop such intelligent systems, artificial intelligence
techniques, as forecasting, can be applied.~Soft computing has been widely
used in real-life applications~\cite{2009:Wu,2012:Taormina}. Furthermore, the estimation of Artificial
Neural Network (ANN) models by using machine learning techniques have been
applied for a wide range of applications, and are also devoted to developing energy systems
~\cite{Ferreira,Karatasou, 2006:energyandbuildings:ruano, 
  2013:energies:zamora}. The problem is that such techniques normally require high computational
resources and historical data, and the traditional training method is based on
batch learning, as for example Back-Propagation (BP) algorithm and its
variants. But for most applications, it could consume from several minutes to some
hours and further the learning parameters must be properly chosen to ensure the
convergence (\emph{i.e.}, learning rate, number of learning epochs, stopping criteria,
\emph{etc.}).~In a batch learning system, when new data are received then it is
performed a retraining jointly with its past data, thus consuming a lot of time
as it is mentioned in~\cite{2003:elsevier:cho,2006:energyandbuildings:ruano}.

Nevertheless, as an alternative, an on-line learning approach could perform the model training with all new incoming data.~In fact, when it is necessary to learn a model from scratch
or to adapt a pre-trained one in a totally unknown scenario, on-line learning
algorithms can be applied successfully~\cite{2013:icann:frandina}.~Thus, we talk with regard to
Stochastic Gradient Descent Back-Propagation (SGBP) as a particular variant of BP for
sequential or on-line learning applications.~Through sequential or on-line
learning methods, the training observations are sequentially presented to the
learning algorithm. Therefore, when new data arrive at any time, they are
observed and learned by the system.~In addition, as soon as the learning procedure is
completed the observations are discarded, without having the necessity to store
too much historical information, that also implies less necessity of additional
physical storage. In conclusion, the learning algorithm has no prior knowledge
about how many training observations will be presented, although it is possible
to produce a better generalization performance at a very fast learning speed [9]
and needs less computing resources that accomplish with our idea of integrating
this technology in a low cost embedded system inside a WSN framework.

The present research group is concerned with regarding the idea of being able to
design new intelligent systems, with few hardware resources, to predict values
of strategic variables related to energy consumption, \emph{i.e.}, low cost and small
predictive systems.~For that purpose, sequential learning algorithms have
demonstrated their feasibility to achieve such objectives.~This also implies
having cheap hardware devices embedding complex artificial intelligence techniques
for forecasting in unknown environments, but also with affordable computing and
economical costs.~As far as we know, it is usual to employ WSN as the monitoring
system that feeds an ANN implemented in a personal computer, as an ANN requires
some complex calculations and that also means using wide data storage.~However, what
it is proposed in this paper is whether or not it is feasible to implement, inside a node of
a WSN, an ANN that performs predictions with an acceptable resolution in its
estimations.~Consequently, in this paper, we present a preliminary
model able to generate low error predictions over short periods of learning
time.~Regarding the innovation of the paper, it has been developed a new on-line
  learning algorithm, based on a BP framework, which is able to preprocess real-time
  continuous input data, incoming in a non-deterministic way, from a wireless
  environment, being also feasible to be implemented in devices with very low
  hardware resources, \emph{i.e.}, with important hardware constraints.

The paper is organized as follows, in Section 2, we describe the framework in
which the present work have been developed, \emph{i.e.}, a wireless sensor network,
describing the hardware of the different nodes and the network topology in order
to slightly describe the experimental setup.~In Section 3, we explain the
approach that has been followed to forecast time series using an on-line
learning paradigm based on Back-Propagation (BP) algorithm for Artificial Neural
Networks.~Section 4 depicts in detail the algorithm developed to be implemented
in a low resources microcontroller as the 8051MCU. Finally, Sections 5 and 6 the
experimental results and the discussion and conclusions explain the present
research and draw some future ideas to continue the present project.

\section{Wireless Sensor Network Architecture}

Basically, a WSN consists of a large number of low-cost, low-power and
multifunctional sensor nodes that are deployed in an environment devoted for monitoring
tasks, but also for controlling as current networks are bidirectional, where
sensor activity can be controlled.~Actually, such sensor nodes, small in size,
are equipped with sensors, embedded microprocessors and radio transceivers, and
therefore they have also capabilities for data processing and communication over
short distances, via a wireless medium, in order to collaborate to accomplish
a common task~\cite{zheng2009wireless}.

A WSN is built of nodes which can vary in number from a few to several hundred
or even thousands, in which each node is connected to one or several
sensors.~Each sensor network node is typically divided into several parts: a
radio transceiver with an internal antenna or connection to an external one,
a microcontroller, an electronic circuit for interfacing with the sensors and an
energy source, usually a battery or an embedded device for energy harvesting~\cite{calvet2014suitability}.~Furthermore, the cost of sensor nodes is also variable, ranging from a few to hundreds of Euros, depending on the complexity of the individual sensor
nodes.~Additionally, size and cost constraints on sensor nodes also result in the corresponding
constraints on resources such as energy, memory, computational speed and
communications bandwidth.~Finally, about the topology of the WSNs, this can vary from a simple star
network to an advanced multi-hop wireless mesh network. In addition, the propagation
technique between the hops of the network can be routing or flooding~\cite{dargie2010fundamentals,sohraby2007wireless}.

Besides sensor networks have the following unique characteristics and constraints, as it
is stated in~\cite{zheng2009wireless}:

\begin{itemize}[noitemsep]
\item Dense Node Deployment. The number of sensor nodes can be of several orders of
  magnitude.
\item Battery-Powered Sensor Nodes. Being in some situations difficult or even
  impossible to change or recharge their batteries.
\item Severe Energy, Computation, and Storage Constraints. Sensor nodes are
  resource limited. This work is focused on this constraint.
\item Self-Configurable. Sensor nodes configure themselves into a communication
  network.
\item Application Specific. A network is usually designed and deployed for a
  specific application.
\item Unreliable Sensor Nodes.~They are prone to physical damages or failures.
\item Frequent Topology Change.~Network topology changes due to node failure,
  damage, addition, energy depletion, or channel fading.
\item No Global Identification. It is usually not possible to build a global
  addressing scheme for a sensor network because it would introduce a high
  overhead for the identification maintenance.
\item Many-to-One Traffic Pattern. In most sensor network applications, the data
  sensed by sensor nodes flow from multiple source sensor nodes to a particular
  sink.
\item Data Redundancy. The data sensed typically have a certain level of
  correlation or redundancy.

\end{itemize}

Furthermore, the characteristics of sensor networks and requirements of different
applications have a decisive impact on the network design objectives in terms of
network capabilities and network performance.~Thus, typically influential design objectives for
sensor networks include the following several aspects: small node size, low node
cost, low power consumption, self-configurability, scalability, adaptability,
reliability, fault tolerance, security, channel utilization and QoS support~\cite{zheng2009wireless}.

Moreover, a typical scheme of a wireless sensor network is composed of a set of nodes that
transmits the information acquired to a sink node.~This one is usually devoted to
collecting and centralizing all the information that comes from the network to a
Personal Computer (PC), in order to store big quantities of data in a persistent
device.~Thus, the information collected can be treated for on-line or later
analysis. But for the purposes of our study, we don't want to dump the
information acquired by the network to a PC. It is desired to use such information, as
it will be described later, to train a neural network, implemented inside a sink node, trying
 to develop an autonomous forecasting system.

\begin{figure}[H]
  \centering
    \includegraphics[width=0.70\textwidth]{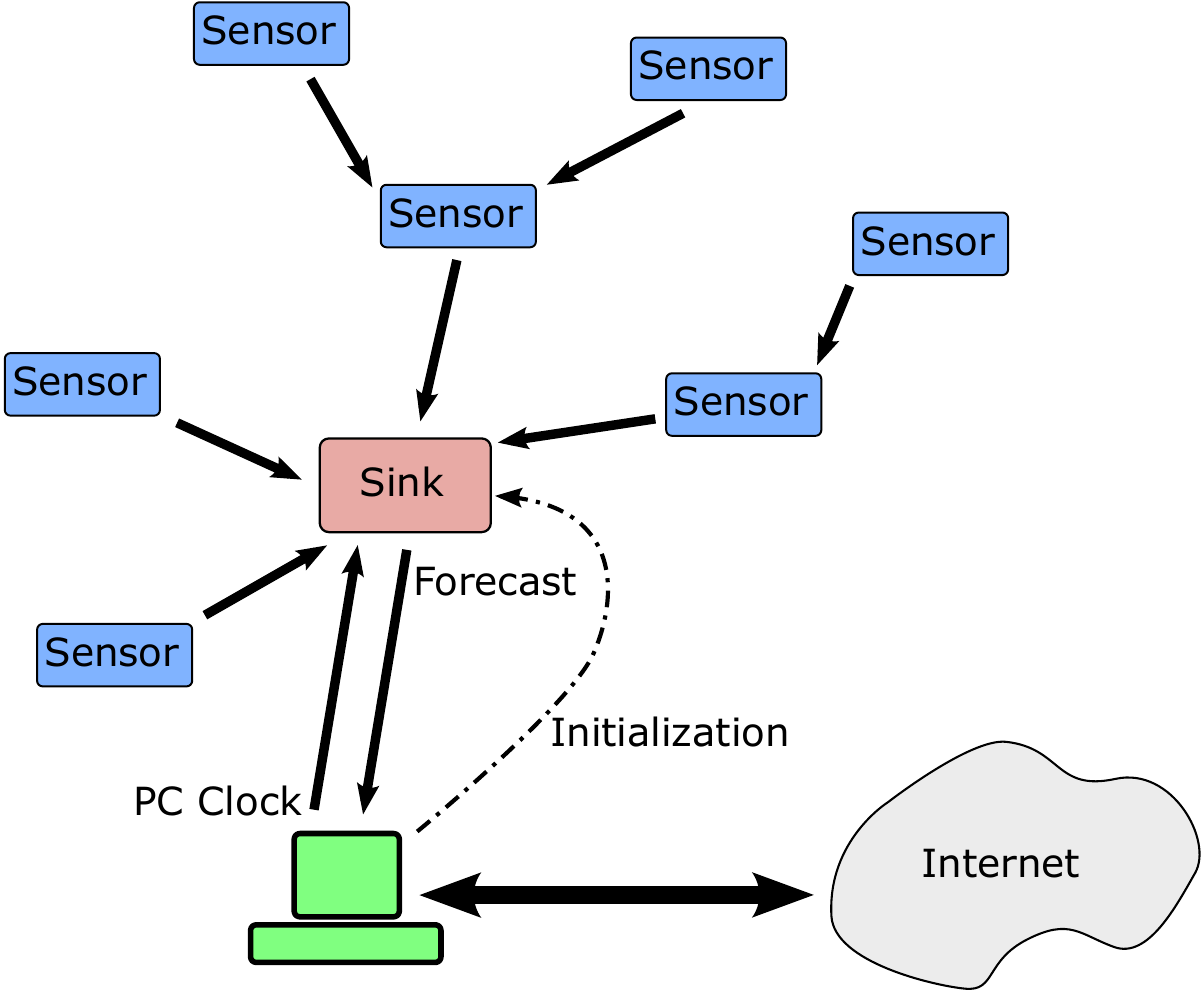}
  \caption{Wireless sensor network scheme.}\label{fig:wireless-sensor-network}
\end{figure}

Figure \ref{fig:wireless-sensor-network} shows the wireless sensor network scheme designed in the present
study.~The network is composed of five nodes, but more nodes can be added in the way the figure displays.~There is a
sink node connected to a PC for configuration and validation purposes and four
sensor nodes that capture the temperature inside a room.~Sensor nodes can work
as repeaters allowing low power transmit modes in order to extend battery
life.

\subsection{Nodes Description}

As mentioned previously, our wireless sensor network consists of two kinds of
nodes, four sensor nodes and one sink node.~Both are based on the same technology,
although, of course, in charge of different tasks.~All nodes are based on the
CC1110F32 microcontroller (Texas Instruments, Dallas, TX, USA)~\cite{readinglow}. The
CC1110F32, is a true low-power sub-1 GHz system-on-chip (SoC) designed for low
power wireless applications.~It combines the excellent performance of the
state-of-the-art RF transceiver CC1101 with an industry-standard enhanced 8051
MCU, up to 32 KB of in-system programmable flash memory and up to 4 KB of Random
Access Memory (RAM), and many other powerful features.~The radio frequency range
can be chosen from: 300--348 MHz, 391--464 MHz and 782--928 MHz.~Its small 6 $\times$ 6 mm
package makes it very suited for applications with size limitations. The
CC1110F32 is highly suited for systems where very low power consumption is
required.~This is ensured by several advanced low-power operating
modes. Additionally, its low power demands (16 mA for transmission at 10 mW and
18 mA for reception) make it suitable for battery-powered systems.

\begin{figure}[H]
  \centering
    \includegraphics[height=0.40\textheight]{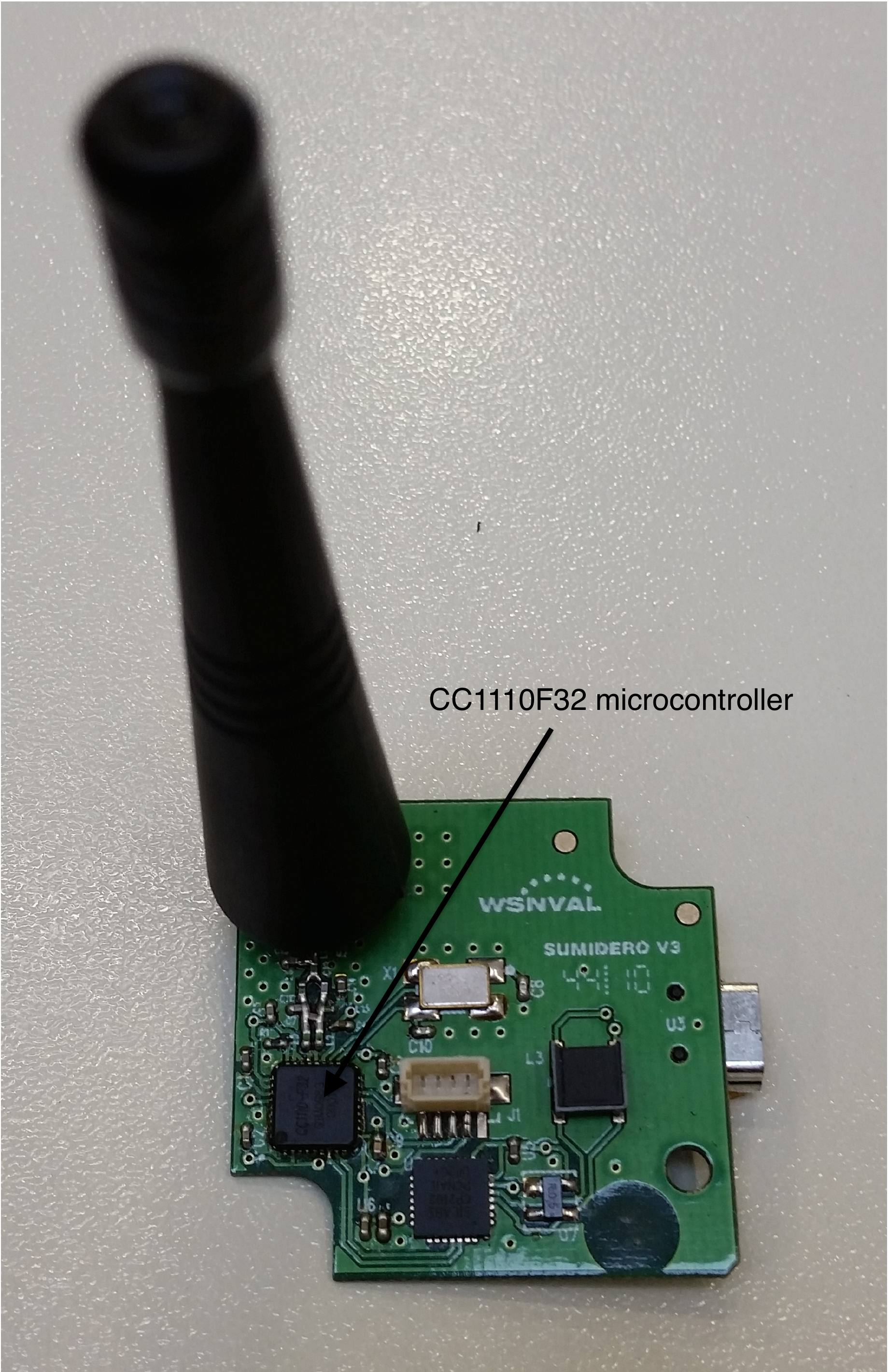}
  \caption{Sink node.}\label{fig:sink}
\end{figure}

\textbf{Sink node} (Figure \ref{fig:sink}): Receives all the wireless information transmitted by the sensor
nodes.~It is switchable to a PC through a USB connection for configuration
purposes, and it works at a speed of 868~Mhz. The USB is also its power
connection, its dimensions are 40~$\times$~40~$\times$~90 mm and it is able to work within a
temperature range of --40 to 85$^\circ$C. 2-FSK, GFSK, MSK, ASK, and OOK modulation
formats are supported and includes a 128-bit AES security coprocessor.~Its power
transmission is 10~mW and high sensitivity of --110 dBm at 1.2 kBaud.~It has an
external exchangeable antenna and programmable data rate up to 500 kBaud.~The
processor also includes one timer of 16 bits and three timers of 8 bits and
on-chip hardware debugging.

\begin{figure}[H]
  \centering
    \includegraphics[width=0.70\textwidth]{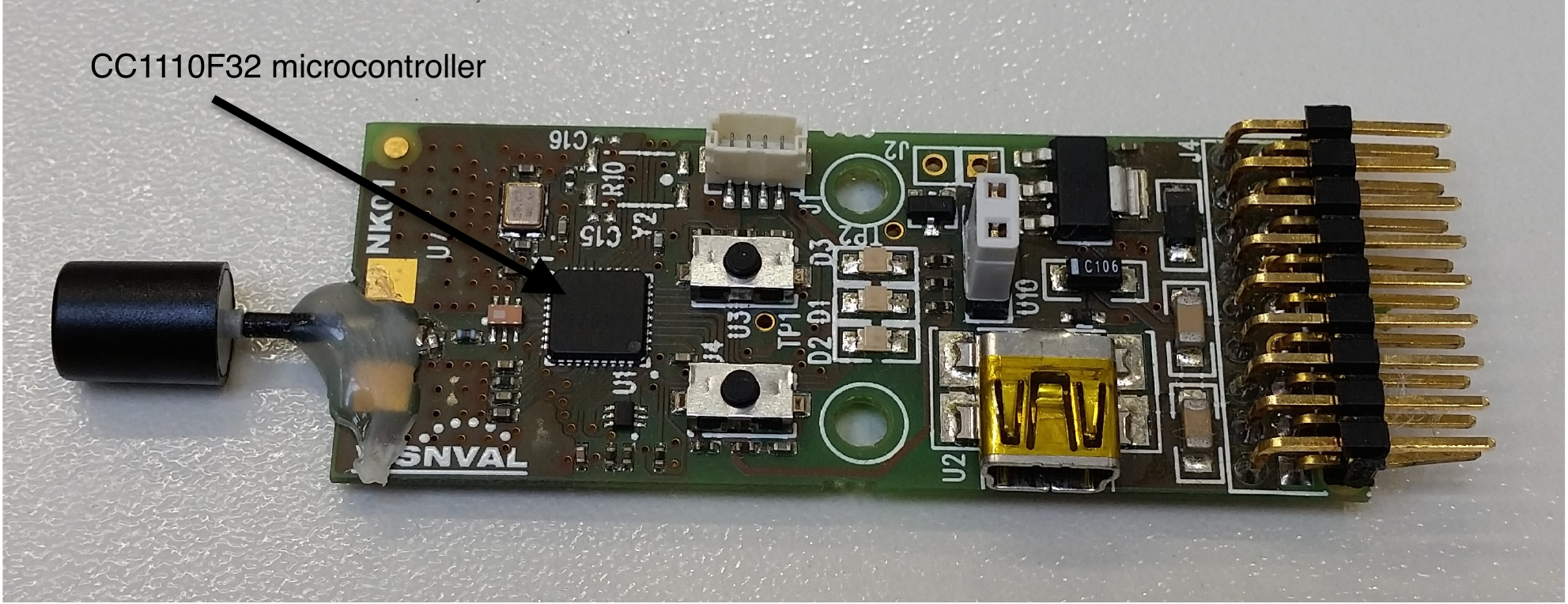}
  \caption{Sensor node.}\label{fig:Sink-node}
\end{figure}

\textbf{Sensor node} (Figure \ref{fig:Sink-node}): There are four sensor nodes, distributed in a room for the present
study, that constantly send the temperature they pick up from the ambient.~All
four have been calibrated with a digital thermometer at the same point and later
have been distributed with similar distances to the sink node.~The sensor node
is based on the same SoC as the sink node and additionally it has a temperature
sensor, three color leds, two buttons (one reset and another for general
purpose) and an input/output expansion connector to install other
sensor/actuators.~Its power supply can be through batteries, USB or electric
power.~In addition, its antenna, which is integrated in the circuit, can reach up to 290~m
without repeaters.~It is possible to connect more sensor nodes, as much as are
needed, depending on the application target.~They autonomously connect themselves to the network and start to send
temperatures~immediately.

The hardware of the sensor and sink nodes have been developed by the company
Wireless Sensor Networks Valencia SL for applications of monitoring and
controlling of industrial systems, domotics, security and alarm systems,
sensorization, telemetry, \emph{etc}. In this study, such nodes have been programmed to
deploy the wireless sensor network and to implement an ANN for on-line learning purposes.

There are other research works developed with more advanced architectures,
as with ARM microcontrollers.~Nevertheless, it is true that the Texas Instruments CC1110F32 MCU
is highly accepted in a lot of areas, and a leader in the market as far as we know,
for Industrial monitoring and control applications.~Moreover, it is also a \emph{true low power SoC}, excellent for WSNs.~If the application is to massively deploy such devices, they are cost competitive and really small.~These devices accomplish with the simplest architecture that makes feasible the implementation of our algorithm.~The SoC has been signaled in the pictures to see the size.

\section{Time Series and Forecasting}

In smart homes is common to measure and monitor environmental variables related
to comfort and energy consumption.~This kind of data are recorded over a
period of time (every second, every minute...).~Studying the behavior of these
variables and forecasting their values in the future allows to manage energy
resources more efficiently.

A collection of data recorded repeatedly over the time represents a time series from
the point of view of an statistical treatment of data. Thus, a time series is a collection of data recorded over a period of time from any
interesting process. They can be formalized as a sequence of scalars from a
variable $x$ obtained as output of the observed process:

\begin{equation}
  x_{t_{1}},\ldots,x_{t_{i-1}},x_{t_{i}},x_{t_{i+1}},\ldots
\end{equation}

There are several methods that are appropriate to model a time series. Time
series data have an internal structure, such as auto-correlation, trend or
seasonal variation and these features should be considered by the modeling and
forecasting methods that can be used.	

Several approaches has been widely employed, such as smoothing and curve fitting
techniques and autoregressive and moving average
models~\cite{chatfield2013analysis,anderson2011statistical,montgomery2011introduction}.~However,
the idea of restricting the storage and computational structure, of our present study, don't allow
us using them.~Other methods framed in regression models have been implemented to
carry out the main purpose of this paper.

In order to simplify as much as possible the forecasting process, in this work,
two different low resource implementable models have been proposed: a linear
model (that is, a Perceptron) and a Multilayer Perceptron (MLP) model with one
hidden layer. Both methods have been implemented and introduced into the sink
node.

Furthermore, with the purpose of comparison and in order to know the behavior of
these two algorithms, a baseline standard method has been developed in a PC.~A
linear model (a perceptron) has been selected and estimated by a standard
method: Bayesian estimation.~A Baseline model approach will represent the results
of a standard model, estimated with more storage prerequisites and with more
complex computational necessities.~It is expected that the results of this
baseline model will be better than those provided by the constrained
models.~However, this comparison will allow to assess, based on the magnitude of
errors obtained, the feasibility of using the algorithms implemented in the
device which has been proposed in this paper.

\subsection{Measurements of Average Model-Performance}

Measurements of average error or model performance are based on statistical
summaries of the differences between each target value vector $\mathbf{y}[t]$ and its predicted
vector $\hat{\mathbf{y}}[t]$.~In general, the most used measure of average model-performance is MAE (Mean Absolute Error):
\begin{equation}
MAE(t=i) = \frac{1}{q} \sum_{z=1}^{q} \vert \hat{y}_{z}[t] -
  y_{z}[t] \vert \,
\end{equation}

The MAE averaged over several time instants in a data set $\mathcal{D}$ will be denoted as MAE$^\star$:
\begin{equation}
MAE^* = \sum_{i} \frac{MAE(i)}{|\mathcal{D}|}
\end{equation}
being  $|\mathcal{D}|$ the number of samples in  $\mathcal{D}$.

\subsection{On-Line Learning in Time Series Forecasting}\label{sec:online}

An on-line learning approach, as the proposed in this work, is a class of
sequential learning paradigm where data frames income on real-time.~The present
work deals with data frames that are not equidistant in time.~This enables the
WSN to be modified at any time, and it allows that the failure of any node at
any moment was not a critical issue.~In Section~\ref{sec:onlinebp} we explain
how the algorithm preprocesses the raw incoming data frames to compute equidistant
and first-order differentiated time series values.~Furthermore, the proposal of
having a low hardware resources constraint for the algorithm implementation, only lets us
to store a short buffer of incoming data, forcing the time series forecasting
model to update its parameters with every new value that arrives.

As far as we know, the
convergence and behavior of on-line algorithms has been studied, by machine learning scientists, as well as the well known Stochastic
Gradient Descent and its variants, and particularly the error Back-Propagation
(BP) algorithm for Artificial Neural Networks~\cite{1988:nature:rumelhart}.  The
on-line BP is stochastic when the training samples are learned in a random
order, probably with replacement.  This randomized update process yields a noisy
gradient computation problem, which is usually overcome by a higher number of
model updates, compared with the batch version of the same BP
algorithm. Therefore, in the standard framework for on-line BP, random traversal
of the dataset yields better results.

However, this work is focused on a straightforward algorithm for real-time time
series learning, that follows an on-line BP strategy but in a sequential
way. Such a problem statement has the advantage of its possible implementation for
low resources devices, but it lacks of the random data traversal of usual
on-line BP approaches, which could be a source of problems in the convergence of
the algorithm. As expected, this problem could be more shocking as more
complex the model is.

A possible refitting, out of the purpose of this work, could be the
implementation of a random skip procedure, that allows for ignoring incoming values
depending on a probability parameter. Such a skip procedure introduces a trade-off
between the stochastic behavior in gradient computation and the number of
samples used to train the model. Additionally, other interesting extensions are
possible for more powerful devices, as a short buffer of input/output samples,
selected randomly from the data stream (known as reservoir sampling), that
performs a model update every $k$ random steps. Even more complicated algorithms
are also possible, where the skip decision rule depends on the behavior of the
model with current data samples.

Different variations of on-line BP algorithm can be taken into consideration for real-time fitting.
  It is true that Least Mean Squares training have problems when input variables are in different scales.
  Thus, Normalized Least Mean Squares~\cite{1995:NLMS:brown} training is proposed to tackle
  with such scale problems. However the developed system adopt first-order differences of input data
  and all input variables belong to the same signal, that is, all input variables have the same scale.
  Other issue of the on-line training algorithm, when exists capture error, is the not convergence
  to a unique value, but to a minimum capture error. This question could be solved by incorporating
  dead-zones in the algorithm as stated in~\cite{1995:NLMS:brown}.

\subsection{Baseline Method: Standard Bayesian Linear Model Estimation}

In linear regression models, observations consist of a response variable in a
vector $\mathbf{y}$ and one or more predictor variables in a matrix
$\mathbf{X}$.~The response vector has $n$ elements, concerning to $n$
observations, so the matrix $\mathbf{X}$ will have $n$ rows and $p$ columns
corresponding to the number of predictors or covariates ($n\geq p+1$ for
non-denerate variance parameter estimation).~Additionally, it is usual to
introduce an intercept parameter in the model, thus a column of
$\mathbf{X}$ is a column completed by number ones.~In~the same way, other exogenous
  variables, which could condition the forecast of the response variable, could be
  introduced as additional columns in matrix $\mathbf{X}$.~Consequently, the
parameters are the regression coefficients, $\mathbf{w}$, and the error variance
of the fitted model, $e^2$. Thus, the linear model could be written in its
matricial form:
\begin{equation}
\mathbf{y}|\mathbf{w}, e^2, \mathbf{X} \sim N(\mathbf{X}\mathbf{w}, e^2 \mathbf{I}_n)
\end{equation}
where $\mathbf{I}_n$ represents the $n \times n$ identity matrix.

Occasionally, more than one future value must be predicted, hence the model receives at each moment as input the $p$ last values in time series and must predict next $q$ values, thus $\mathbf{y}$ and $\mathbf{w}$ becomes in a $n
\times q$ and $p \times q$ matrices $\mathbf{Y}$ and $\mathbf{W}$. In this
framework, at time $t$ it is available one new element of $p$ past values in the time series ($[x_{t-p+1},...,x_{t-1},x_{t}]$) that was considered as covariates in the simple linear model
 to forecast $q$ further values $(x_{t+1},...,x_{t+q})=(y_1,....,y_q)$. In consequence, it has been built a simple linear model
for each prediction:
\begin{equation}
\mathbf{Y}_{.i}|\mathbf{W}_{.i}, e_{i}^{2}, \mathbf{X} \sim N(\mathbf{X}\mathbf{W}_{.i}, e_{i}^{2} \mathbf{I}) \;\; i=1,...,q
\label{eq:lin}
\end{equation}
where $\mathbf{M}_{.i}$ denotes $i$-column at any $\mathbf{M}$ matrix.

This forecasting process needs $n+1$ observations to start to generate predictions ($\mathbf{X}$ represents a $(n+1) \times p$ matrix and $\mathbf{Y}$ a $(n+1) \times q$ matrix) and $q$ linear models must be estimated (one for each prediction) in each time step.

Some assumptions must be considered about the classical linear model, such as
$\mathbf{X}$ should be full rank (no collinearity among predictors),
exogenous predictors and not auto-correlated errors.~Using this sort of models
to represent time series, that uses lagged predictors to incorporate feedback
over the time, means that some of these assumptions are violated. Autoregressive processes comparable to this one
introduce violations of classical linear model assumptions that lead to biased
parameters estimations.~Nevertheless, it could be improved, considering more complex covariance structures.~As it has been mentioned before, a random skip procedure could solve some of these problems.~However, in this case, the primary goal of our work has been focused on building an on-line estimation model process.~Moreover, it has to be able to perform predictions with an acceptable resolution in its estimation, and with low computing and
storage~resources.

A Bayesian framework~\cite{2013:crc:gelman} provide a natural way to perform an on-line
estimation.~Such methods make it possible to incorporate scientific hypothesis
or prior information based on previous data, by means of the prior
distribution.~Nevertheless, in the absence of prior information, a Bayesian estimation of the parameters is made in an objective context, with objective prior
distributions for them, that are estimated with information provided by
the data.~In addition, when prior information is available or some scientific hypothesis have been
assumed, a Bayesian estimation could be made in a subjective framework,
incorporating prior information into parameter prior distributions.

In the context of this work, the first step is a Bayesian parameter estimation,
made by means of a non-informative prior distribution for the
parameters.~When the first parameter estimation is available then the estimated model is
used to generate the predictions.~The predictive distribution, $ \hat{\mathbf{Y}}$,
given a new set of predictors $\mathbf{X}_{p}$ has mean:
\begin{equation}
 \hat{\mathbf{Y}}_{.i}=\mathbf{X}_{p} \hat{\mathbf{W}}[0]_{.i} \;\;i=1,...,q
\end{equation}
where $\hat{\mathbf{W}}[0]$ is calculated with first $n+1$ data elements available $\mathbf{X}$ and $\mathbf{Y}$ as:
\begin{eqnarray}
\hat{\mathbf{W}}[0]_{.i} &=& (\mathbf{X}^{\intercal}\mathbf{X})^{-1} \mathbf{X}^{\intercal} \mathbf{Y}_{.i}
\end{eqnarray}


Therefore, it demonstrates that it is necessary to solve matrix products and inverse matrices
with dimension $(p \times p)$. Inversion matrices could be avoided employing,
for example, $QR$ decomposition in $\mathbf{X}$ matrix, but in terms of
computational and storage cost remains as an expensive process.

After this first step, the system has prior information available, that must be
incorporated in a future parameters estimation.~The way to introduce previous
information, with new data to improve such parameters estimation, is by using
informative prior distribution.~In the context of this work, linear regression
with an informative prior distribution was used. At time $t$, previous
estimation on parameter assessment was employed in prior distribution of parameters
and new data was utilized to re-estimate model parameters.~Furthermore, if at time $t$ the
last parameter estimation is $\hat{\mathbf{W}}[t-1]$ and we have new data
$\mathbf{X}[t]$ and $\mathbf{Y}[t]$, the new appraisal at this point of time
can be computed by treating the prior as additional data points, and then
weighting their contribution to the new estimation \cite{2013:crc:gelman}.~To
perform the computations, for each prediction value $\mathbf{Y}[t]_{.i} \;\;
(i=1,...,q)$ it is necessary to construct a new vector of observations
$\mathbf{Y}^{*}_{.i}$ with new data and last parameters assessment, and
predictor matrix $\mathbf{X}^{*}$, and weight matrix $\bm{\Sigma}$ based on previous
variance parameters estimation as follows (more details can be read at the~Appendix):
\begin{eqnarray}
\mathbf{Y}^{*}_{.i} &=& \left[
      \begin{array}{c}
        \mathbf{Y}[t]_{.i} \\
        \hat{\mathbf{W}}[t-1]_{.i} \\
      \end{array}
     \right]\\
\mathbf{X}^{*} &=& \left[
      \begin{array}{c}
        \mathbf{X}[t] \\
        \mathbf{I}_{p} \\
      \end{array}
     \right]
\end{eqnarray}
where $\mathbf{I}_{p}$ represents a $p \times p$ identity matrix. The
new parameters appraisal at time t could be written as:
\begin{eqnarray}
\hat{\mathbf{W}}[t]_{.i} &=& (\mathbf{X}^{*\intercal} \bm{\Sigma}^{-1} \mathbf{X}^{*})^{-1}
\mathbf{X}^{*\intercal} \bm{\Sigma}^{-1} \mathbf{Y}^{*}_{.i} \;\;i=1,...,q
\end{eqnarray}

Thus, the predictive distribution, $\hat{\mathbf{Y}}$, given a new set of
predictors $\mathbf{X}_{p}$ has as a mean:
\begin{equation}
 \hat{\mathbf{Y}}_{.i}=\mathbf{X}_{p} \hat{\mathbf{W}}[t]_{.i}\;\;i=1,...,q
\end{equation}

Complexity in second and posterior steps in the assessment process is higher than
the first step.~The number of matrix inversions and computational complexity has
increased now.

Consequently, assuming a simple linear model, with the limitations mentioned before that
this type of models have to model dynamic processes, it represents a computational and
storage costs too high to be implemented in a device with low hardware resources, as we are concerned in this
paper.~The Bayesian standard linear model parameters estimation is a simple
process but with high computational resource requirements.~Thus, it is necessary to
solve some inverses matrices, whose cost is too expensive in the context of this
paper.

In conclusion, this model has been considered as baseline model to compare the
results of both algorithms related with the two ANN models that have been
implemented in the sink node.

\section{Sequential On-Line Back-Propagation Algorithm for Devices with Low Resources}\label{sec:onlinebp}

This section describes the considered implementation of an on-line version of BP
algorithm~\cite{1988:nature:rumelhart} for devices with very few memory and
computing resources, as the 8051 microcontroller included in the sink node.~The
BP algorithm is able to train any kind of ANN. As it was stated at
Section~\ref{sec:online}, the present research proposes the utilization of a
sequential version of on-line BP, and compares two different models with
previous stated baseline: a linear model (that is, a perceptron) as shown in
Equation~(\ref{eq:perceptron}) and equivalent to the model of
Equation~(\ref{eq:lin}), and a MLP with one hidden layer as shown in
Equation~(\ref{eq:mlp}):
\begin{eqnarray}
  \hat{\mathbf{y}} &=& \mathbf{W}_1 \cdot \mathbf{x} + \mathbf{b}_1 \label{eq:perceptron}\\
  \hat{\mathbf{y}} &=& \mathbf{W}_2 \cdot s(\mathbf{W}_1 \cdot \mathbf{x} + \mathbf{b}_1) + \mathbf{b}_2 \label{eq:mlp}
\end{eqnarray}
being $\mathbf{W}_j$ weight matrices, $\mathbf{b}_j$ bias vectors,
$\mathbf{x}$ the input and $\hat{\mathbf{y}}$ the output of the ANN.~The
Appendix~\ref{sec:bp} describes the derivation of BP algorithm to train any kind
of ANN, from perceptrons to MLPs with any number of hidden layers.
Moreover, the input vector $\mathbf{x}$ can also be extended with other exogenous
variables that could influence the output response variable.~BP is a kind of first order
gradient descent algorithm, therefore it needs the computation of partial
derivatives over $\mathbf{W}_j$ and $\mathbf{b}_j$.~BP algorithm has been
chosen because of its simplicity, it depends in algebra operations as dot
products, matrix-vector products and component-wise products. The proposed MLP
has the logistic activation function in the hidden layer and the activation function
can be implemented by means of an exponential one.~Such operations are
available by hardware and/or software libraries in 8051 microcontroller, and
the memory requirements to implement these operations depend on the complexity
of the ANN developed:

\begin{itemize}[noitemsep]

\item In the case of the perceptron, it needs a weights matrix $\mathbf{W}_{1}$
  of size $p \times q$, a bias vector with $q$ elements, input and output
  vectors with size $p$ and $q$, and an output gradients vector with $q$
  elements. The BP algorithm would need $p \times q + p + 3q$ real numbers in
  memory to work with the perceptron, lets consider $p = q = 8$. Thus, it would need $96$
  real numbers.

\item In case of MLP with one hidden layer of length $h$, it needs two
  weights matrices $\mathbf{W}_{1}$ and $\mathbf{W}_2$. The first one with size $p
  \times h$, the second one with $h \times q$; two bias vector, the first one with $h$
  and the second one with $q$ elements; the input, hidden and output vectors, with
  size $p$, $h$ and $q$ respectively; and the output and hidden gradients vector
  with $q$ and $h$ elements also correspondingly. The BP algorithm needs $p \times h + h
  \times q + 2p + 3h + 2q$ real numbers, lets consider $p = h = q = 8$. Thus, it would need
  $184$ real numbers in memory.

\end{itemize}

Using $32$ bit precision for real numbers, the memory resources of BP algorithm
for one hidden layer MLP when $p=h=q=8$ are $184 \cdot 4 = 736$ bytes.

A BP algorithm has been introduced into the source code that has been implemented for the sink
node. This algorithm has been split in three parts and basically computes
differences in mean temperature every quarter (15 min), and handles these data
to learn the ANN model. The memory requirements of this algorithm and its characteristics in detail will be described
in the next sections. For the sensor nodes, the source code is not described as they only send data to the sink node, the temperature.

\subsection{Main Loop Algorithm}

The experimental setup consisted of the equidistant placement of four sensor
nodes in the living room of a solar house.  They send continuously the
temperature to a sink node that is in charge of predicting the mean temperature
for the next hours.  The sink node is connected to a PC, mainly for some
configurations and power reasons, and it was placed at a small work place in the
living room.~The sink node sends the temperature predictions to the home's
central control.~It is an application that was developed by the present group
to monitor and control all the energy systems for the purpose of the Solar
Decathlon competition \cite{abcde}. Its name is CAES
(Computer-Aided Energy Saving System) system~\cite{2013:energies:zamora}.

The main loop procedure implemented in the sink device is shown in
Algorithm~\ref{alg:main}.~Before the main loop, the sink node starts with some
initializations, as the board support package, the minimal radio frequency (RF)
interface and the rest of issues corresponding to the wireless network.~To do
that, it has been compiled the SimpliciTI protocol from Texas Instruments, that
is a simple low-power RF network protocol aimed at small RF networks.~All of those
aspects correspond to the \FUNCNAME{initializeWSN}() function.~After that, the
weights initialization, required by the ANN, are established through the
function \FUNCNAME{initializeWeights}().~This can be done in a random way, thus
the ANN starts from scratch or it can be also done reading such weights from the
computer, in this way the system starts with a pre-trained ANN. It was decided
to define the parameters randomly.

The core of the main loop first receives from each sensor node a data frame
($v$), which corresponds to the \FUNCNAME{waitForDataFrame}() procedure.~It
includes the temperature and some information to identify the node
sender.~Sensor nodes send the temperature constantly, and as they are
equidistant placed in a medium size room, the difference in temperature among
them is minimum, thus it can be considered redundant for the present application.~Although the most important for our study is to be able to process continuous messages coming from different wireless nodes to perform an on-line
learning as stated before.~When a data frame is received,
its time stamp ($t$) is collected through the function \FUNCNAME{askForCurrentTimeStamp}()
and then it is called the subroutine
\FUNCNAME{processSampleOnLine}$(v,t)$. Such a function receives a time stamp and a
temperature value, and computes averages of quarters (15~min) feeding these
averages to the ANN. Such averages are computed as the integration between
previous (time, temperature) pair and the current one.~When a quarter is ready,
it is given to the ANN as described~later.

This main loop has negligible memory requirements; it only uses static variables
to pass the data between the different subroutines.~The memory consumption due
to the wireless communication protocol will be ignored in this paper, it is not
the focus of the presented work.~Nevertheless, depending on the data frames
nature and the implementation of \FUNCNAME{processForecastOutput}$(o)$ function,
the whole algorithm can be used for different applications.~In the present work,
the goal is to predict the indoor temperature, and thus every data frame is a
temperature value given by a wireless node.

\begin{algorithm}[H]
  \caption{Main procedure for sink node.}\label{alg:main}
  \begin{algorithmic}[1]
    \STATE{\FUNCNAME{initializeWSN}()}
    \STATE{\FUNCNAME{initializeWeights}()}
    \WHILE{\TRUE}{%
      \STATE{$v = $ \FUNCNAME{waitForDataFrame}$()$}
      \STATE{$t = $ \FUNCNAME{askForCurrentTimeStamp}$()$}
      \STATE{$o = $ \FUNCNAME{processSampleOnLine}$(v,t)$}
      \STATE{\FUNCNAME{processForecastOutput}$(o)$}
    }\ENDWHILE
  \end{algorithmic}
\end{algorithm}
\vspace{-15pt}
\subsection{On-Line Sample Processing}

The subroutine \FUNCNAME{processSampleOnLine}$(v,t)$ is displayed at
Algorithm~\ref{alg:process}.~This subroutine receives a pair of temperature
frame $v$ and the timestamp $t$ in seconds, and executes one iteration of the
on-line algorithm.~This algorithm is
responsible for the computation of mean values of temperature every $Q$
seconds.~Once a mean value is available (which happens every $Q$ seconds), the
algorithm calls the procedure \FUNCNAME{TrainAndForecast}$(v_q)$ which computes
the first order differentiation of these means and adjusts the model weights.~At
the presented work, $Q=900$~s (15~min, a quarter) because it is a
reasonable value for temperatures.

\begin{figure}[H]
  \centering
  \includegraphics[width=0.68\textwidth]{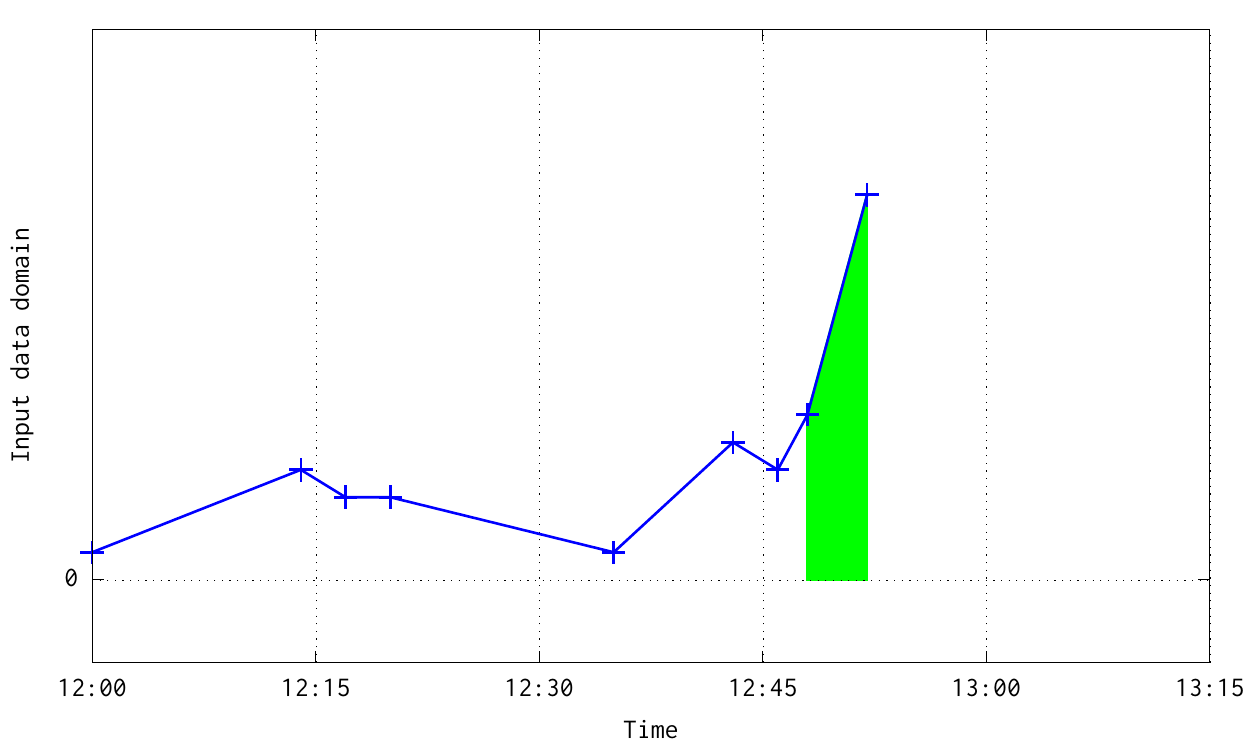}
  \caption{Illustration about the integration process of previous and current input
    data pairs.}\label{fig:case1}

\end{figure}

During temperature mean computation, due to the non deterministic nature of the
WSN, the data frames maybe are not equidistant in time.~This algorithm considers this
issue implementing an procedure of aggregation for the mean temperature
computation~(For this aggregation, it is assumed that every node is
  sensing similar temperatures, so their readings can be mixed-up without
  problems.).~Every pair of consecutive input data, which belongs to the same time
quarter, are integrated and aggregated to an accumulator variable (see
line~\ref{alg:process:aggregation} at Algorithm~\ref{alg:process}) following
this equation:
\begin{equation}
  \mathbb{A}(t_{i},v_{i},t_{i+1},v_{i+1}) = \frac{t_{i+1} - t_{i}}{Q} \cdot \frac{1}{2} \left( v_{i} + v_{i+1} \right)
  \label{eq:integrate}
\end{equation}
where $\langle t_i, v_i \rangle$ and $\langle t_{i+1}, v_{i+1}
\rangle$ are the pair of input data.~The Figure~\ref{fig:case1} shows a
graphical illustration of this process for the case when the pair of data are into
the same quarter.
\begin{algorithm}[H]
  \begin{algorithmic}[1]

    \caption{\FUNCNAME{processSampleOnLine}($v$,$t$)}\label{alg:process}

    \REQUIRE{$v,t$ are \emph{real numbers} with data value and
      timestamp in seconds. $v^\prime, t^\prime, q_t^\prime, v_q$ are \emph{static
        variables} initialized with invalid values, and used by the algorithm
      to store the data value, the time and quarter number in previous function call,
      and in the end to aggregate the data value for current time quarter (the
      algorithm computes the mean every quarter.)  The \emph{constant} $Q=900$
      is the number of seconds in a quarter. The algorithm interpolates quarter
      mean values when they are missing, up to a maximum of $M = 4$
      quarters.}

    \ENSURE{A prediction vector, in case it is possible to be computed,
      otherwise it returns a \FUNCNAME{NULL} value.}

    \STATE{$q_t = \floor{\displaystyle{\frac{t}{Q}}}$}
    \IF{$t^\prime$ is not a valid value}{%
      \STATE{$v_q \leftarrow v \cdot \displaystyle{ \frac{(t \mod Q)}{Q} }$}
      \COMMENT{Initializes accumulated data for aggregation}
    }\ELSE{%
      \IF[Quarter change limit exceeded]{$q_t - q_t^\prime > M$}{%
        \STATE{\FUNCNAME{reset}$()$}\label{alg:process:reset}
      }\ELSE{%
        \STATE{$\langle m,b \rangle = \mathbb{L}(t^\prime, v^\prime, t, v)$}
        \COMMENT{Interpolates line $v_i = m \cdot t_i + b$ using Equation~(\ref{eq:line})}
        \STATE{$t_i = Q \cdot (q_t^\prime + 1)$}
        \WHILE{$t_i \leq t$}{\label{alg:process:while}
          \STATE{$v_{t_i} = t_i \cdot m + b$}
          \STATE{$v_q \leftarrow v_q + \mathbb{A}(t^\prime, v^\prime, t_i, v_{t_i})$}
          \COMMENT{Aggregates data change following Equation~(\ref{eq:integrate})}
          \STATE{$\mathbf{o}_{t_i} = \FUNCNAME{trainAndForecast}(v_q)$}
          \COMMENT{Last $\mathbf{o}_{t_i}$ is stored to be returned at function end}
          \STATE{$v_q \leftarrow 0.0; v^\prime \leftarrow v_{t_i}; t^\prime \leftarrow t_i$}
          \STATE{$t_i \leftarrow t_i + Q$}
        }\ENDWHILE
        \IF{$t^\prime < t$}{%
          \STATE{$v_q \leftarrow v_q + \mathbb{A}(t^\prime, v^\prime, t, v)$}\label{alg:process:aggregation}
          \COMMENT{Aggregates last data change following Equation~(\ref{eq:integrate})}
        }\ENDIF
      }\ENDIF
    }\ENDIF
    \STATE{$t^\prime \leftarrow t; v^\prime \leftarrow v; q_t^\prime \leftarrow q_t$}
    \RETURN{$\mathbf{o}_{t_i}$ if available, otherwise \FUNCNAME{null}}
  \end{algorithmic}
\end{algorithm}

\begin{figure}[H]
  \centering
  \includegraphics[width=0.68\textwidth]{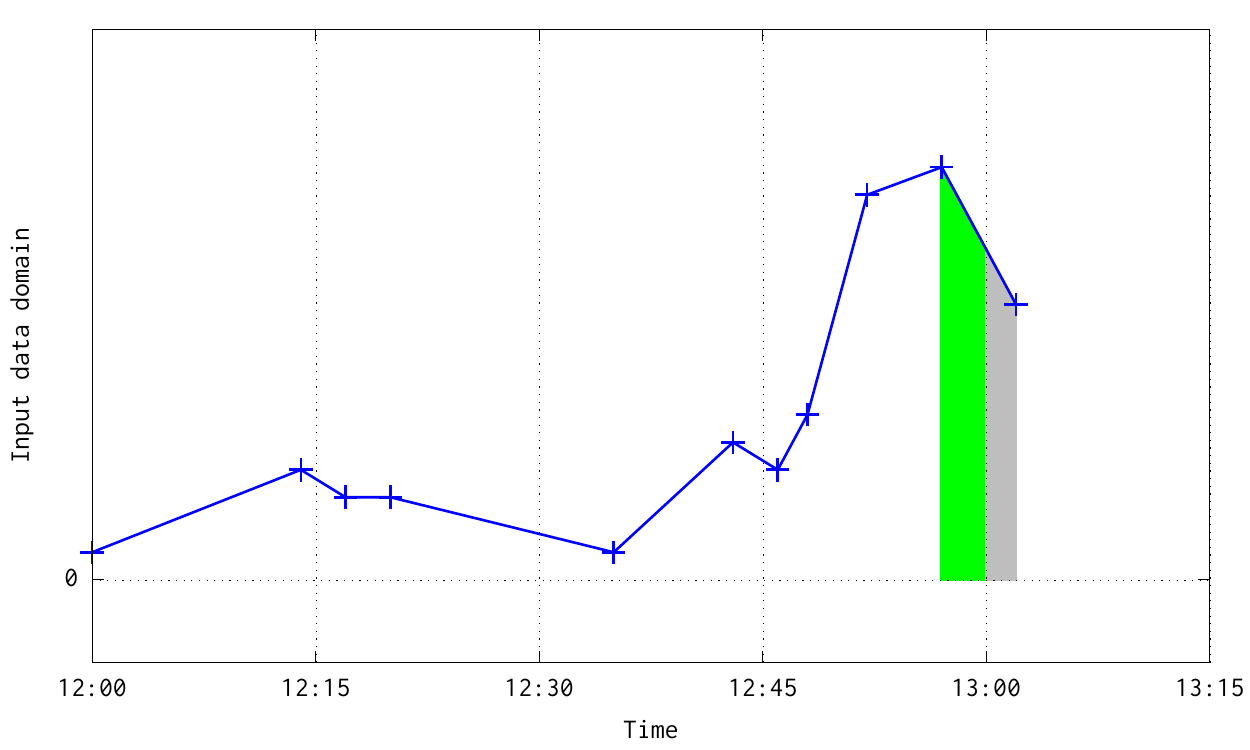}
  \caption{Illustration about the integration process of previous and current input
    data pairs when both are in different but in consecutive quarters. In green
    color has been shown the interpolation computed until 13:00, ending a
    quarter mean value computation. In gray color has been depicted the
    interpolation that would be aggregated to the quarter ending at
    13:15.}\label{fig:case2}
\end{figure}

\begin{figure}[H]
  \centering
  \includegraphics[width=0.68\textwidth]{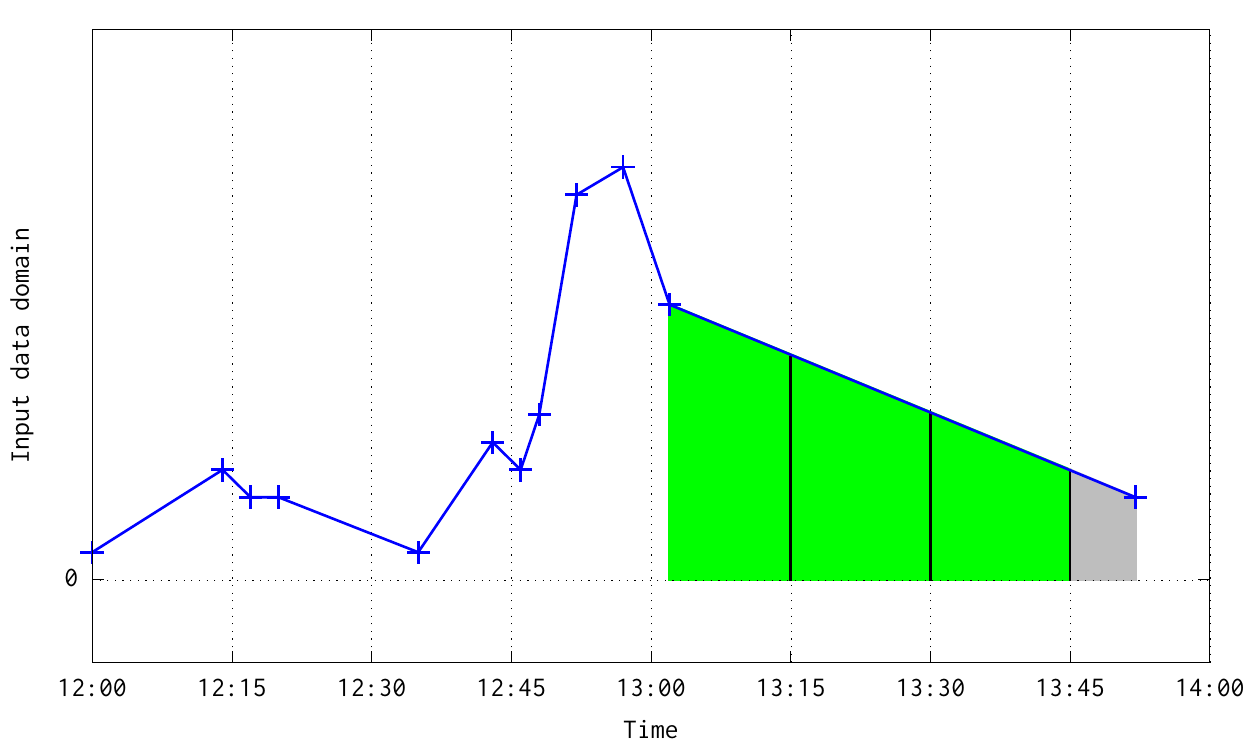}
  \caption{Illustration about the interpolation and integration of previous and current
    input data pairs when they are in non consecutive quarters. In green
    color has been depicted the interpolation of the computed quarters that end at
    13:15, 13:30 and 13:45. In gray color has been shown the interpolation
    aggregated to the following quarter that would end at
    14:00.}\label{fig:case3}
\end{figure}

This aggregation has two boundary cases: when two consecutive pairs are in a
different but consecutive time quarters (see Figure~\ref{fig:case2}); or when two
consecutive pairs are in different and non consecutive quarters due to the lost
of a large number of frames (see Figure~\ref{fig:case3}).~In both situations, besides
the case when a quarter is fully processed, are solved at the loop at
line~\ref{alg:process:while}.~Before the loop, it is interpolated a line
equation which follows the temperature slope between the pair of available input
data, computed with the next slope-intercept linear equation:
\begin{equation}
  \mathbb{L}(t_{i},v_{i},t_{i+1},v_{i+1}) = \langle m = \frac{v_{i+1} - v_{i}}{t_{i+1} - v_{i}}, ~~~ b = v_{i} - m \cdot t_{i} \rangle \,
  \label{eq:line}
\end{equation}

The loop begins at the start point, and traverses the line segment by
$Q$ seconds increments, computing the mean temperature of every possible quarter
between both input pairs. When a quarter is completed, its mean value is given
to the subroutine \FUNCNAME{trainAndForecast}$(v_q)$.~In the extreme case of
losing a huge number of frames, the whole system is reset at
line~\ref{alg:process:reset} of Algorithm~\ref{alg:process}, starting the
process again but without initializing the model weights (This reset
  procedure also initializes the static variables of Algorithm~\ref{alg:train}.).

The memory requirement for subroutine \FUNCNAME{processSampleOnLine}$(v,t)$ is again
negligible.~It uses only a few static variables to aggregate the data values
and/or interpolate the lost quarter values.~For temperature, data values are in
$^\circ$C.

 \vspace{-12pt}

\subsection{On-Line Training and Forecast using Back-Propagation}

The last subroutine is depicted at Algorithm~\ref{alg:train}.~It computes
the difference between two consecutive quarter means, and stores them into an
auxiliary circular buffer $B$ with length $p+q$.~A~counter $k$ is used to
control the number of items in the buffer, controlling if it is possible to
produce a forecast, and/or when it is feasible to perform one training step of
the model.~In particular, the forecast condition is true when the buffer counter
$k$ is greater or equal to the model's input size $p$. On the other hand, the
train condition is true, when the buffer counter $k$ is greater or equal to the
sum of model's input $p$ and output $q$ sizes.

The algorithm uses a static variable $v^\prime$ where the value given at
previous function call is stored.~Thus, in the first call, the \verb+if+ statement
at line~\ref{alg:train:valid} is not executed.~In the following calls, the first
order differentiation is computed at line~\ref{alg:train:diff} and the counter
of elements is increased by one unit.~The training condition is checked at
line~\ref{alg:train:train}, and in case of success the input/output mapping is
taken from the circular buffer $B$ and the model is updated following BP
Equations~(\ref{eq:forward1})--(\ref{eq:update2}).~The forecast condition is
checked at line~\ref{alg:train:forecast}, and in case of success the input is
taken from the last $p$ items of the buffer $B$, then the forecast is produced
following Equations~(\ref{eq:forward1})--(\ref{eq:forward2}).~Finally, the
output vector is dedifferentiated by computing the cumulative sum of the model
output and adding it up with all the vector components and the input data value of
current quarter.

Note that the reset call at line~\ref{alg:process:reset} of
Algorithm~\ref{alg:process} also initializes the static variables of this
one.~This algorithm has more critical memory requirements, due to the circular buffer
$B$.~The length of such buffer is $p+q$, and in the experimentation these values
are $p=q=8$, therefore, this algorithm needs $16$ real numbers, that using $32$
bit real numbers, corresponds to $64$ bytes.~Thus the total memory consumption
needed by the whole algorithm (BP + on-line control) in the worst case adds up
to $64 + 736 = 800$ bytes.

\begin{algorithm}[H]
  \caption{\FUNCNAME{trainAndForecast}($v_q$)}\label{alg:train}
  \begin{algorithmic}[1]

    \REQUIRE{$v_q$ is a \emph{real number} with the value at current
      quarter. $v_q^\prime$ and $k = 0$ are \emph{static variables}, the first
      one stores a quarter value given in previous function call and it is
      initialized with an invalid value. The second one is a counter initialized
      with $0$ needed to access the auxiliary buffer $B$.  $p$, $q$ are
      \emph{constants} with the input size, output size and buffer size respectively.  $B$ is
      a \emph{static circular buffer} with $p+q$ length. For simplicity, $B$ is
      indexed with any integer value $i \geq 0$, assuming that $i \mod (p+q)$ is
      needed to access to valid positions. $B$ buffer stores the value
      difference between two time quarters. The forecast starts when the counter
      $k$ is $p$, and the training starts when the counter $k$ is at least
      $p+q$.  The weight matrices $\mathbf{W}_{j}$ and $\mathbf{b}_{j}$
      (both randomly initialized at start), activation vectors $\mathbf{h}_{j}$
      and error gradients $\bm{\updelta}_{j}$ are \emph{global variables},
      used by \FUNCNAME{forward}, \FUNCNAME{backprop} and \FUNCNAME{update}
      functions. $\upeta_0$ is the initial learning rate, $\upgamma$ is the learning
      rate decay value, and $\upepsilon$ is the weight decay, all of them are
      \emph{constants}.~These last three parameters have been set after a grid
      search optimization is done, with different values depending on the
      model. $\upalpha=0$ is another \emph{static variable} which is the number of
      performed learning iterations.}

    \ENSURE{The algorithm trains the model using functions
      \FUNCNAME{forward}, \FUNCNAME{backprop} and \FUNCNAME{update}, and returns
      the forecast at current time quarter, or \FUNCNAME{NULL} if it cannot be
      computed.}

    \IF{ $v_q^\prime$ is a valid value }{\label{alg:train:valid}
      \STATE{$B[k] = v_q - v_q^\prime$}\label{alg:train:diff}
      \STATE{$k \leftarrow k + 1$}
      \IF[Train when buffer $B$ is full]{$k >= p + q$}{\label{alg:train:train}
        \STATE{$\mathbf{x} = B[(k - p - q):(k - q - 1)]$}
        \STATE{$\mathbf{y} = B[(k - q):(k - 1)]$}
        \STATE{$\hat{\mathbf{y}} = \FUNCNAME{forward}(\mathbf{x})$}
        \COMMENT{Following Equations~(\ref{eq:forward1})--(\ref{eq:forward2})}
        \STATE{$\FUNCNAME{backprop}(\hat{\mathbf{y}}, \mathbf{y})$}
        \COMMENT{Following Equations~(\ref{eq:backprop1})--(\ref{eq:backprop2})}
        \STATE{$\upeta = \displaystyle{\frac{\upeta_0}{ (1 + \upalpha \cdot \upeta_0)^{\upgamma}}}$}
        \STATE{$\FUNCNAME{update}(\upeta, \upepsilon)$}
        \COMMENT{Following Equations~(\ref{eq:update1})--(\ref{eq:update2})}
        \STATE{$\upalpha \leftarrow \upalpha + 1$}
      }\ENDIF
      \IF[Compute forecast]{$k >= p$}{\label{alg:train:forecast}
        \STATE{$\mathbf{x} = B[(k - p):(k - 1)]$}
        \STATE{$\hat{\mathbf{y}} = \FUNCNAME{forward}(\mathbf{x})$}
        \COMMENT{Following Equations~(\ref{eq:forward1})--(\ref{eq:forward2})}
        \STATE{$\mathbf{o} = \FUNCNAME{cumSum}(\hat{\mathbf{y}}) + v_q$}
        \COMMENT{The outputs vector $\hat{\mathbf{y}}$ is dedifferentiated by
          computing the cumulative sum of the vector and adding-up
          current quarter value to all vector components}\label{alg:train:cumsum}
      }\ENDIF
    }\ENDIF
    \STATE{$v_q^\prime \leftarrow v_q$}
    \RETURN{$\mathbf{o}$ if available, otherwise \FUNCNAME{null}}
  \end{algorithmic}
\end{algorithm}

In this algorithm the training and forecasting procedures are delayed by $q$
iterations ($q$ time quarters). Thus the algorithm produces forecasts using a model
trained with data of $q$ quarters in the past.

\subsection{Last Remarks}

\begin{figure}[H]
  \centering
  \includegraphics[width=0.68\textwidth]{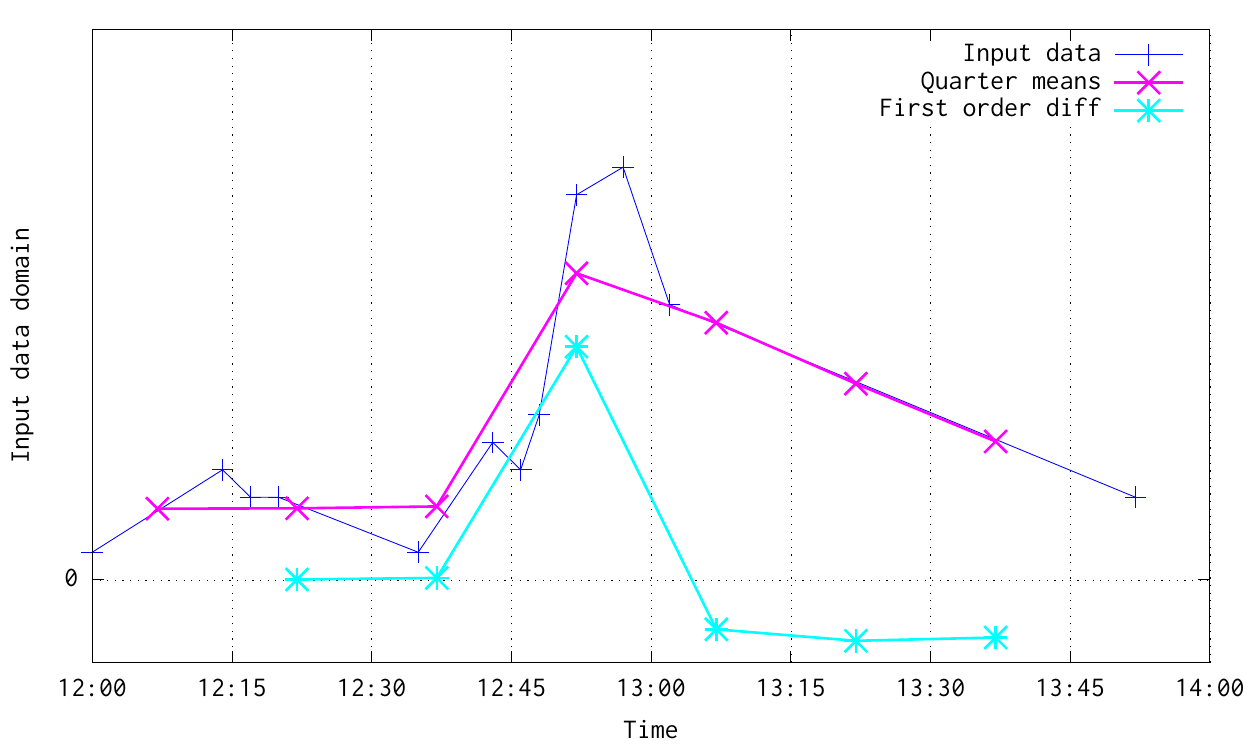}
  \caption{Illustration of input data, means and first order differentiated
    result.}\label{fig:means}
\end{figure}

As has been stated before, following an on-line method, this algorithm trains
an ANN model for time series forecasting. In order to improve the model
performance, the input data is aggregated and filtered by computing its mean
every $Q$ seconds. After that, first order differentiation of the data is
computed, and the model is trained to predict the differentiated series.
Figure~\ref{fig:means} shows an illustration of input data, its transformation
into means every $Q$ seconds, and the differentiated series used to train the
system. Mean and differentiated series have been plotted in the mid-point of
every $Q$ window.

This algorithm has been proposed to receive as input a fixed number of
  delayed past values. This limits the model ability to learn and forecast
  variables which are conditioned to exogenous data (For instance, in
    case of indoor air temperature prediction, HVAC operations and other related
    variables should be taken into account to ensure good model performance.).
  Nevertheless, it is straightforward to extend the algorithm to receive
  additional data as input, which would be passed directly to the model as
  covariates.

\section{Results and Discussion}

For the study of the proposed algorithm, two case studies have been
completed.~The first one is a simulation using artificially generated data, with a
very large number of iterations in order to have a baseline to compare model behavior and accuracy estimate.~The second one is a real application for indoor temperature forecasting, using the dataset SML2010 DataSet~\cite{sml2010} at UCI
Machine Learning repository~\cite{Bache+Lichman:2013}.~The SML2010 DataSet has been created by the ESAI research team (the authors of the present paper) at the Universidad CEU Cardenal Herrera, monitoring the data captured from a solar
domotic house that participated in the Solar Decathlon Europe 2010, a world competition of energy efficiency.
Such a dataset has been employed as it is tidy, \emph{i.e.}, it has been cleaned and structured, and it is ready for analysis.

\subsection{A Simulation Study}

\begin{figure}[H]
  \centering
    \includegraphics[width=0.7\textwidth]{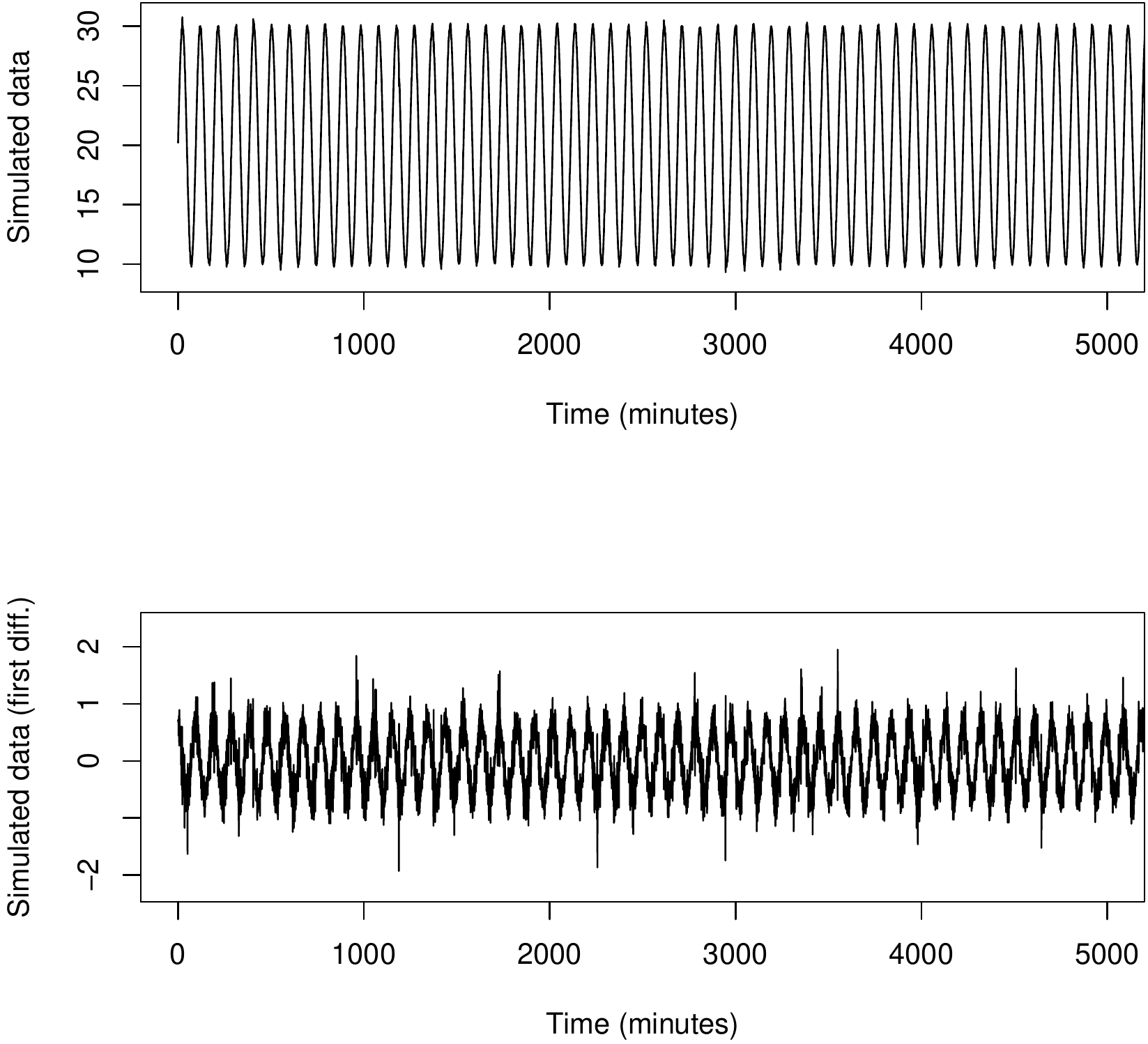}
  \caption{First 5000 simulated time series data and first differences.}\label{fig:time-series-dataSin}
\end{figure}

A simulation study has been carried out to deeply explore the algorithm behavior.~The
simulated dataset contains $10^6$ data, based on a sinus function in different
time points along $8500$ h, with centigrade degree values between $10$ and $30$ (more
variability than real temperature values).~Furthermore, the distance between two consecutive
values was randomly taken from a range of $[20,40]$ s.~Sin values have also been randomly
modified with noise in a range of $[-1.5,1.5]$. The original dataset was on-line
preprocessed by the mean in order to obtain one value of temperature every $15$
minutes.~Moreover, in order to increase model generalization, first differences on
preprocessed data were calculated, obtaining the final dataset that was
modeled. Note that the proposed on-line BP algorithm computes the mean and first order differences on-the-fly during the training process.
This dataset is shown in Figure~\ref{fig:time-series-dataSin}.

\begin{figure}[H]
 \centering
  \includegraphics[width=0.7\columnwidth]{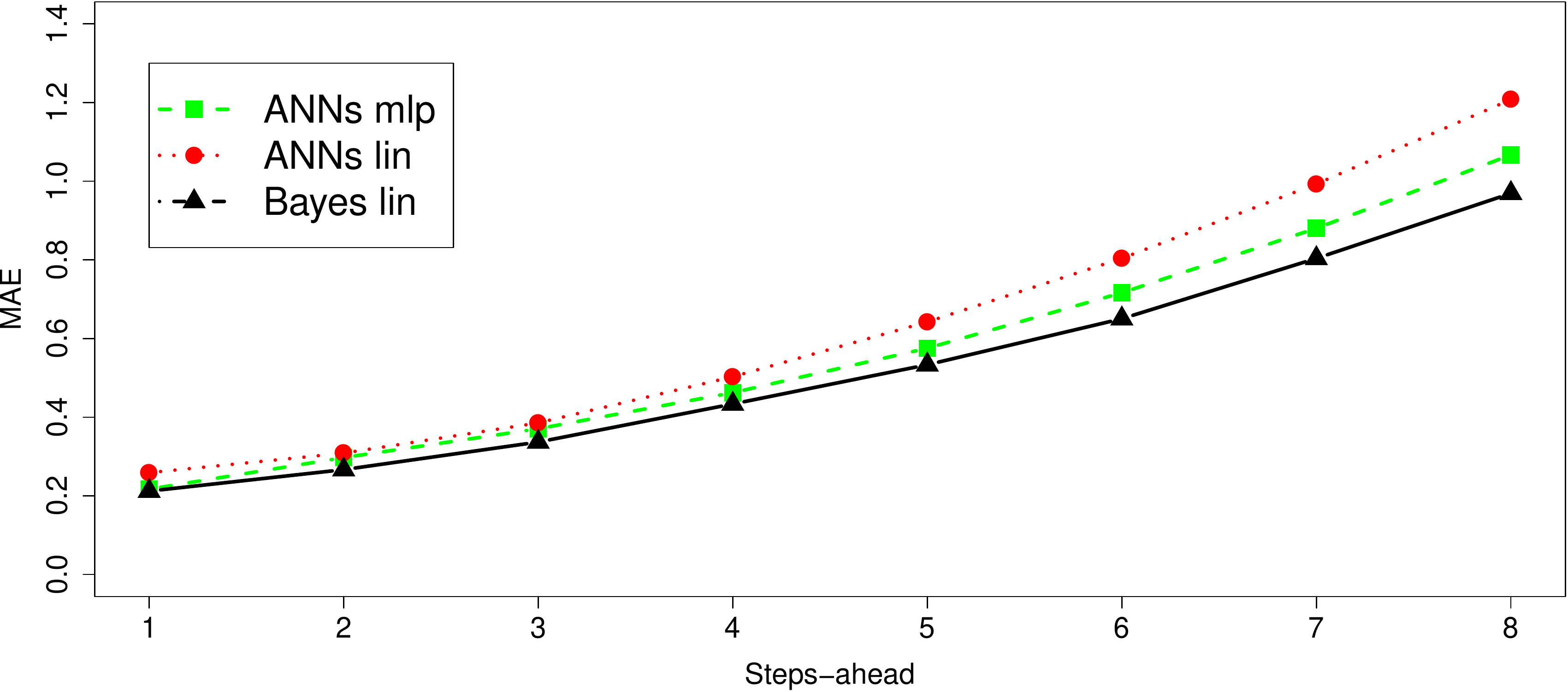}
  \caption{MAE of each step-ahead for last 15,000 for sin data.}\label{fig:maesinlag}
\end{figure}

\begin{figure}[H]
 \centering
  \includegraphics[width=0.7\columnwidth]{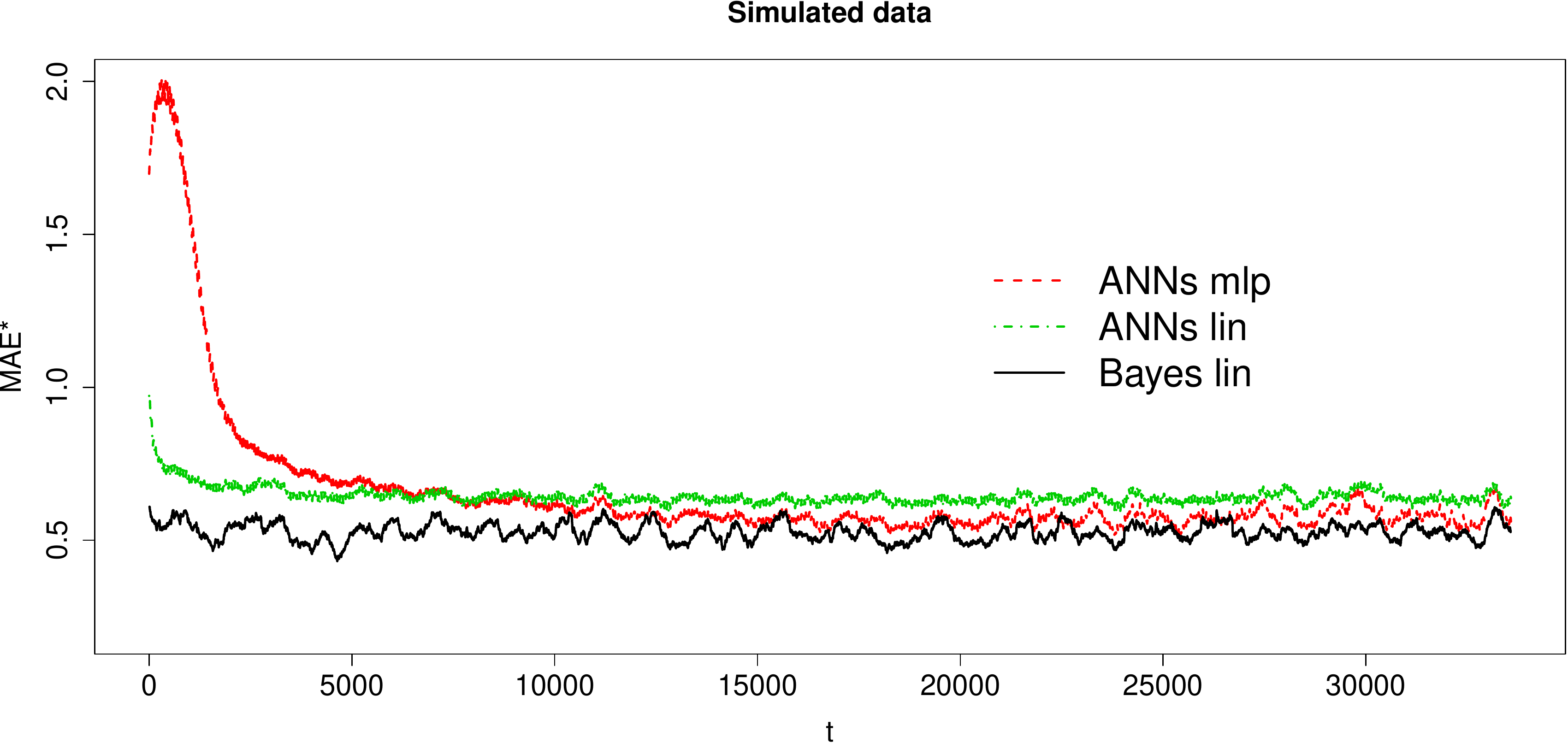}
  \caption{Smoothed MAE$^\star$---sin data.\label{fig:maesin}}
\end{figure}

Figure~\ref{fig:maesinlag} and \ref{fig:maesin} show the MAE and MAE$^\star$
  behavior. First, the Figure~\ref{fig:maesinlag} illustrates how the errors
  evolve with the number of steps-ahead.~The first 15,000 observations have not
  been considered in this calculation, because they belong to the period in
  which the convergence of the algorithms had not been reached.~Second, the Figure~\ref{fig:maesin} shows the smoothed MAE$^\star$ behavior, calculated
by 10 values window length to avoid randomness noise over the time and the
Table~\ref{tab:comparisonSin} depicts the MAE summarized obtained from the
dataset for the baseline Bayesian model and the implementable ANN models. The
input/output structure was defined choosing 8 values as input and 8 future
values as output. Thus, the model receives at each moment as input, the last
eight values in time series and it must predict next eight values (next two
hours by steps of 15 min).

Figure~\ref{fig:maesin} and Table~\ref{tab:comparisonSin} demonstrate the comparison
between the Bayesian baseline and the performance obtained by the linear and MLP
models.~It is possible to observe how the Bayesian baseline is able to obtain
low errors almost from the first iteration, because of its better utilization of the
available information. The structure employed in this baseline model is higher than the other two methods.
The baseline model operates in each iteration with a $n \times p$ matrix $\mathbf{X}$, with $n=p+1$, and a vector $n \times 1$ vector $\mathbf{Y}_{.i}$.~However, ANN models handles only a $1 \times p$ vector $\mathbf{x}$ and a vector $1 \times 1$ vector $\mathbf{y}$.~Regarding ANN models it is observed that the linear model learns
faster than MLP, nevertheless MLP achieves better error in the long-term. As expected, the
more complex the model is the more difficult the learning process, but it achieves similar results
to the Bayesian baseline in the long run.

\begin{table}[H]
  \caption{Comparison between Bayesian baseline and ANNs models. Both
    methods has a $p=8$ inputs and $q=8$ outputs. The MLP has a hidden layer
    with $h=8$ neurons.}
  \label{tab:comparisonSin}
  \centering
  \small
  \begin{tabular}{ccccccc}
    \toprule
    \textbf{Method} & \textbf{Min.} & \textbf{Q1} & \textbf{Q2} & \textbf{Mean} & \textbf{Q3 }& \textbf{Max.} \\
    \midrule
    Baseline (Bayesian standard) & 0.047 &  0.248 & 0.442  & 0.528 &  0.720 &  4.227\\

    Lin & 0.036  & 0.368 &  0.632 &  0.648 &  0.877 &  2.991 \\
    MLP & 0.046 &  0.323 &  0.553 &  0.662 &  0.871  & 3.708 \\
    \bottomrule
  \end{tabular}
\end{table}
\subsection{Real Application: Temperature Forecasting in a Solar House}

\begin{figure}[H]
 \centering
  \begin{tabular}{cc}
  \includegraphics[width=0.35\columnwidth]{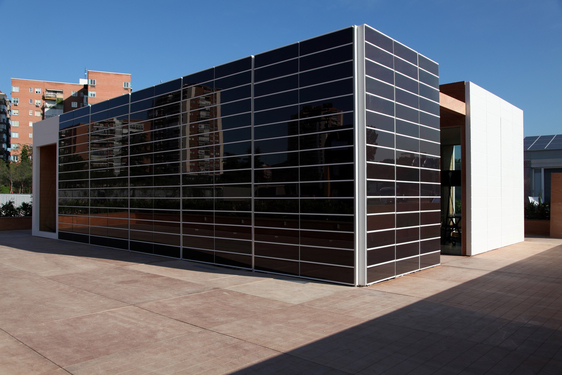}
	\includegraphics[width=0.35\columnwidth]{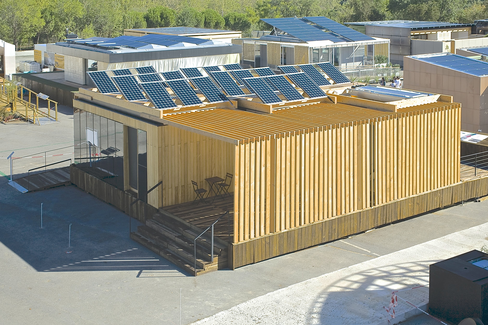}\\
	\end{tabular}
  \caption{Solar-powered houses: SMLhouse (\textbf{left}) and SMLsystem (\textbf{right}).\label{fig:solarhouses}}
\end{figure}

As was mentioned before, the School of Technical Sciences at the University CEU Cardenal Herrera (CEU-UCH)
participated in 2010 and 2012 at Solar Decathlon Europe competition building two solar-powered houses known
as SMLhouse and SMLSystem respectively (Figure~\ref{fig:solarhouses}).~One of the technologies integrated in that houses was a monitoring
system developed to collect all the data related with energy consumption and other variables as indoor/outdoor
temperature, CO2, \emph{etc.}~A tidy database was created and it has been exploited to develop different
ANN models for prediction purposes in previous research projects.

\begin{figure}[H]
  \centering
    \includegraphics[width=0.65\textwidth]{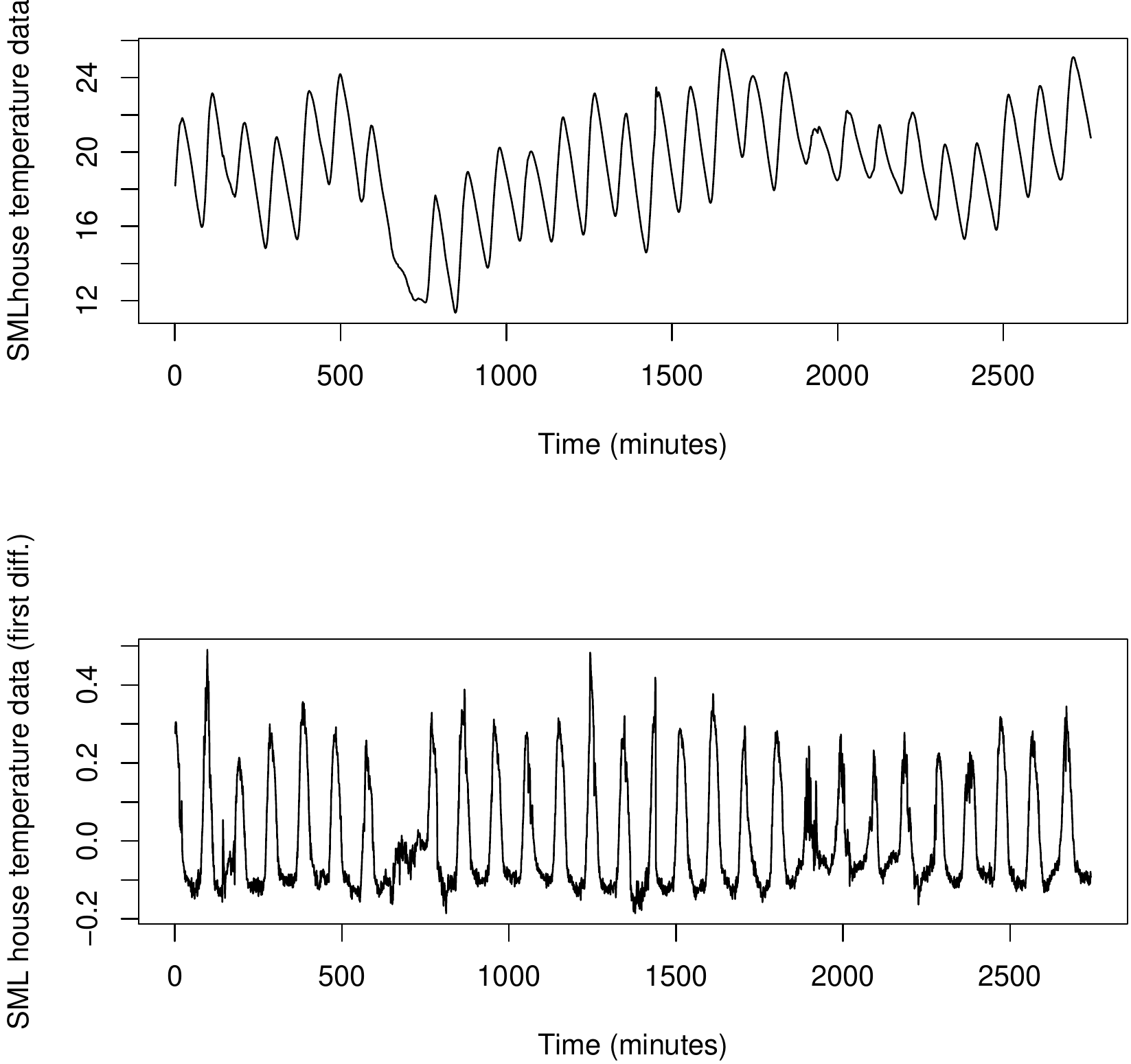}
  \caption{SMLhouse time series data.}\label{fig:time-series-dataSML}
\end{figure}

As stated before, a tidy dataset, with $673$ h ($28$ days) of real
temperature data, recorded every $15$ min is available ($2688$ equally
spaced temperature data) and it has been utilized to study the performance of
the overall system.~This dataset and its first difference is shown in
Figure~\ref{fig:time-series-dataSML}.

\begin{figure}[H]
  \centering
  \includegraphics[width=0.7\columnwidth]{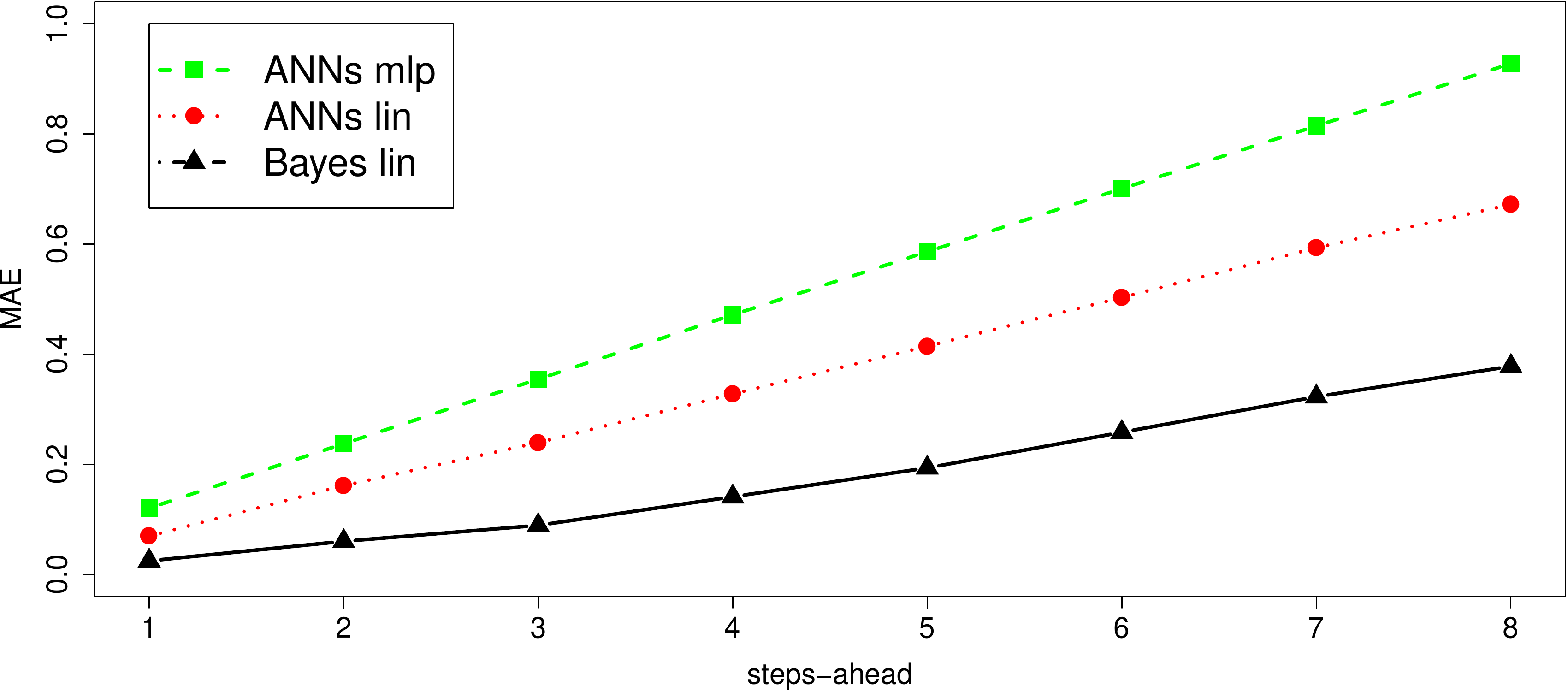}
  \caption{MAE of each step-ahead for SML house forecasts.} \label{fig:maesmllag}
\end{figure}

\begin{figure}[H]
  \centering
  \includegraphics[width=0.7\columnwidth]{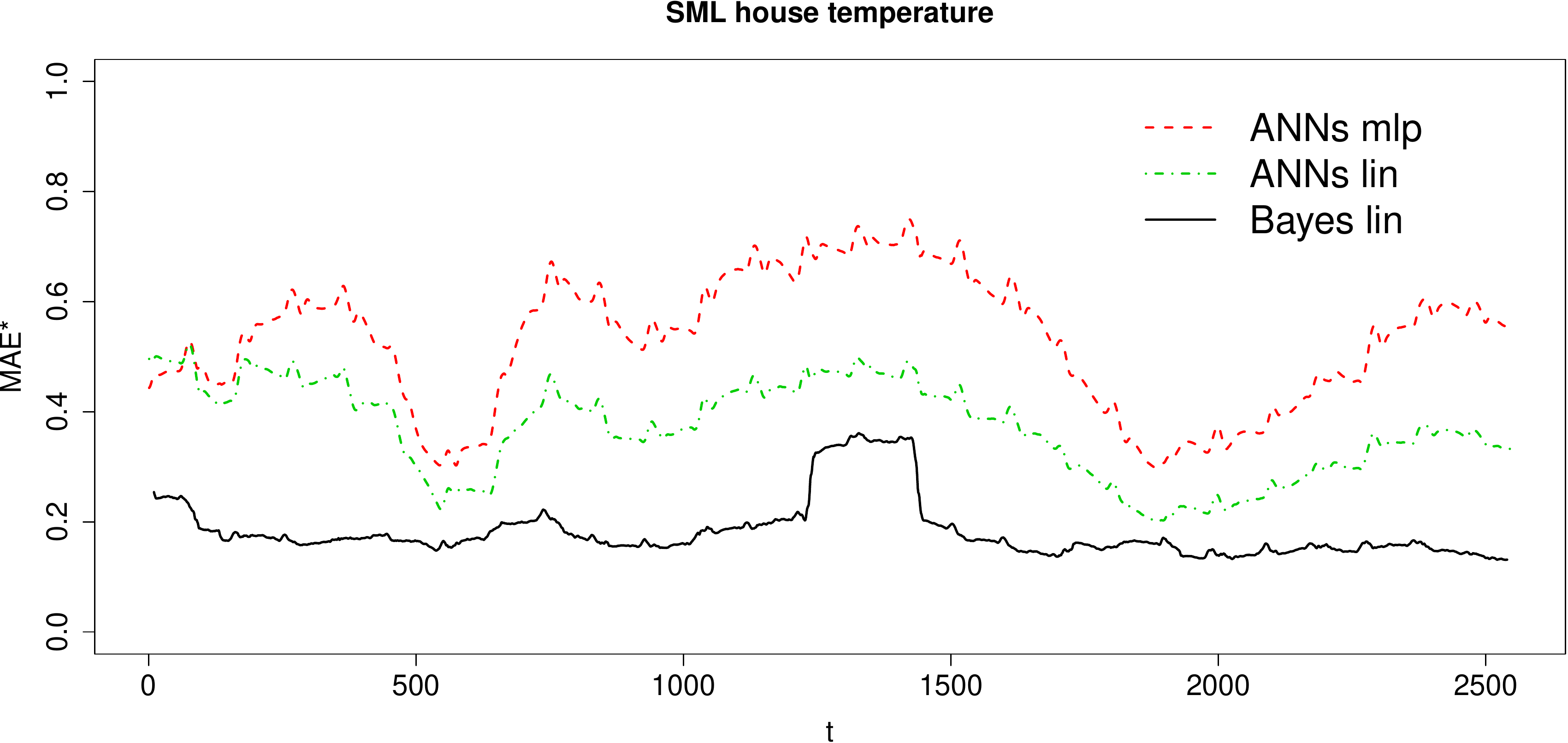}
  \caption{Smoothed MAE$^\star$---SML house. \label{fig:maesml}}
\end{figure}

In the same way as the study of the simulated data, Figures
  ~\ref{fig:maesmllag} and \ref{fig:maesml} show de MAE and MAE$^\star$ behavior.~Figure~\ref{fig:maesmllag} illustrates how the errors evolve with the number
  of steps-ahead and Figure~\ref{fig:maesml} visualizes the smoothed MAE$^\star$
  behavior, calculated by 10 values window length to avoid randomness noise over
  the time.~Table~\ref{tab:comparisonSML} shows the summarized MAE obtained with
  the dataset for the baseline Bayesian model and the implementable ANN
  models.~The input/output structure was defined choosing 8 values as input
and 8 future values as output.~Furthermore, the model receives at each moment as
input the last eight values in time series and must predict the next eight
ones (next two hours by step of 15~min), as it has been described for the
baseline before.~Because of the short observation period, compared to
the observation period of simulated data, ANNs models do not converge to the
baseline model results.~However, as it has been demonstrated in the simulation
study, it is expected that as the time evolves, the errors tend to be
equal. Anyway, although it can be observed how the Bayesian baseline is able to
obtain low errors again, nevertheless the absolute differences between this
method and ANN models is negligible in practical terms.

The experimental results obtained seem very promising, as we were able to implement
a complex algorithm in a simple hardware device to predict time series accurately.
In previous research, published in the journal Energy and Buildings~\cite{2014:EandB:zamora}
this kind of algorithms, in analogous situations but with additional variables, were able to obtain
superior predictions and converge to low errors in lower periods of time.
Such issue makes us to consider dealing with the possibility of using the present algorithm and the same ideas
to develop a new one that will employ similar variables, as the previous work,
but to be implemented in hardware devices with higher resources, keeping the restrictions of using
cheap and small-sized microcontrollers.

\begin{table}[H]
  \caption{Comparison between Bayesian baseline and ANNs models. Both methods
    has a $p=8$ inputs and $q=8$ outputs.}
  \label{tab:comparisonSML}
  \centering
  \small
  \begin{tabular}{ccccccc}
    \toprule
    \textbf{Method} & \textbf{Min.} & \textbf{Q1} & \textbf{Q2 }& \textbf{Mean} & \textbf{Q3} & \textbf{Max.} \\
\midrule
    Baseline (Bayesian standard) & 0.008 &  0.087 &  0.141 &  0.184 &  0.222 &  4.295\\
    Lin ANN & 0.011 &  0.213 &  0.309 &  0.373 &  0.432 &  1.978 \\
    MLP ANN & 0.013 &  0.233 &  0.466 &  0.527 &  0.734 &  2.109 \\
  \bottomrule
  \end{tabular}
\end{table}


\section{Conclusions}

This paper describes how to implement an artificial neural network in a low cost
system-on-chip to develop an autonomous intelligent wireless sensor network. The model
is able to produce predictions of temperature coming from four sensor nodes
based on an on-line learning approach.~The idea behind this project is to evaluate if
it is possible to develop, in a system with very few resources as the MCU 8051,
the necessary source code to integrate a neural network that learns a time
series in an on-line strategy,\emph{ i.e.},\ without any historical database.~It is
obvious that all kind of problems cannot be afforded with this
technology/approach.~Nevertheless, on some physical measurements, represented as a stationary
time series, it is possible to apply an on-line learning paradigm.~And it makes attainable to generate
predictions of time series in feasible temporal windows, in the present case study
in few days.~That means that it is conceivable to place a low cost and small intelligent system
in a totally unknown environment to learn its dynamics very rapid and with a
wireless technology.

The on-line learning approach is suboptimal in terms of the model accuracy.~The
proposed algorithm is able to learn accurate forecast models, but as stated at
Section~\ref{sec:online}, learning in a sequential way could harm the model
learning.~Stochastic gradient descent algorithms are based on random sampling of
training dataset, and minor changes in the proposed algorithm can be performed
to allow some class of random sampling in the real-time data stream.~The
implementation of these ideas in more complex devices would allow to increase
model and of course algorithm complexity, in time and space, improving the
experimental results shown in this work.~An idea for the future is to scale this project to
architectures more complex but starting with an efficient algorithm as a baseline.

Finally, some issues would arise when using the proposed models to predict the
  temperature of a room where HVAC system is operating. Air temperature would be
  affected by HVAC operations, among other exogenous variables, thus forecasting
  only with a fixed number of past delayed values would be not enough. However,
  this problem can be tackled by extending the model with additional input information
  regarding the operations performed by the HVAC system and other exogenous
  variables that determine the room temperature.
\newpage
\acknowledgments{Acknowledgments}

This work has been supported by
Consolidaci\'on de indicadores CEU-UCH2014-15 program of the
Vicerrectorado de Investigaci\'on at Universidad CEU-Cardenal Herrera.~The authors wants to acknowledge the valuable comments and suggestions provided by the reviewers, and the valuable work done by the edition team.


\authorcontributions{Author Contributions}

Juan Pardo has been contributed with WSN programming and nodes configuration, besides
leading this research. Francisco Zamora-Mart\'inez has been contributed with
design and development of on-line learning algorithm and he has run experiments
with ANNs. Paloma Botella-Rocamora has been contributed with Bayesian baseline
experiments besides discussions and ideas. All of them have contributed to
the writing, discussions and preparation of the manuscript.


\appendix
\setcounter{equation}{0}
\renewcommand\theequation{A\arabic{equation}}

\section{Standard Bayesian Linear Model Estimation}

This section describes some aspects of the standard bayesian estimation for regression linear model parameters in which our work is based.

\subsection{Non-Informative Prior Distributions}

The linear model could be written:

\begin{equation}
\mathbf{y}|\mathbf{w}, e^2, \mathbf{X} \sim N(\mathbf{X}\mathbf{w}, e^2 \mathbf{I})
\end{equation}

The non-informative prior distribution most commonly used for
linear regression is:
\begin{equation}
P(\mathbf{w}|e^2) \propto \displaystyle \frac{1}{e^2}
\end{equation}

The posterior distribution of $\mathbf{w}$ given $e^2$ and the marginal
posterior distribution of $e^2$ could be obtained analytically as follows:
\begin{eqnarray}
\mathbf{w}|e^2,Y &\sim& Normal(\mathbf{w}, V_{\mathbf{w}} e^2)\\
e^2|Y &\sim& Inverse \chi^2(n-p,s^2)
\end{eqnarray}

The parameters estimation of these distributions could be calculated as follows:
\begin{eqnarray}
\hat{\mathbf{w}} &=& (X^\intercal X)^{-1} X^\intercal  \mathbf{y}\\
V_{\mathbf{w}} &=& (X^\intercal X)^{-1}\\
s^2 &=& \displaystyle \frac{(\mathbf{y}-X \hat{\mathbf{w}})^\intercal (y-X \hat{\mathbf{w}})}{n-p}
\end{eqnarray}

The symbol $\intercal$ denotes matrix transposition. It would be needed $n>p$ for carry
out parameters estimation and $n=p+1$ has been considered in the process estimation ($p \times (p+1)$ data).

If $q$ must be predicted in each time step, $q$ linear models must be estimated. Thus, $q$ estimations of vector $\mathbf{w}$ ($\mathbf{w}_{i}$ for $i=1,...q$) could be represented as $\mathbf{W}$ matrix, that contains in each of its $q$ columns the estimated parameter vector $\mathbf{w}_i$ for each prediction.
$$
\mathbf{W}_{.i}=w_i
$$

In the same way, vector of predictions $\mathbf{y}$ becomes in a matrix $\mathbf{Y}$ with $q$ columns. Vector column $\mathbf{W}_{.i}$ corresponds with parameter vector in the linear model with response vector $\mathbf{Y}_{.i}$. Each of these models has the same predictor matrix $\mathbf{X}$.

At time $t=p+p+1$ a first estimation $\mathbf{W}[0]$ is available and can be used to make
predictions at time $t+1$. The predictive distribution, $\hat{\mathbf{Y}}$,
given a new set of predictors $\mathbf{X}_{p}$ has mean:
\begin{equation}
 \mathbf{\hat{Y}}_{.i}=\mathbf{X}_{p} \hat{\mathbf{W}}[0]_{.i} \;\;i=1,...,q
\end{equation}

With variance for this estimation:
\begin{equation}
var(\hat{\mathbf{Y}}_{.i}|\mathbf{e}_{.i}^2,Y)=(I+\mathbf{X}_{p} \mathbf{V}_{w} \mathbf{X}_{p}^{^\intercal }) \mathbf{e}_{.i}^{2}
\end{equation}

It is necessary to solve matrix products and inverse matrices with dimension $(p
\times p)$.

\subsection{Informative Prior Distribution}

Furthermore, in the following time points $t$, if last parameter estimation is $\hat{\mathbf{W}}[t-1]$, $s^2[t-1]$ and $V_{\mathbf{W}}[t-1]$ and we have new data
$\mathbf{X}[t]$ and $\mathbf{Y}[t]$, new parameters estimation at this point of time
can be computed by treating the prior as additional data points, and then
weighting their contribution to the new estimation \cite{2013:crc:gelman}. To
perform the computations, for each prediction value $\mathbf{Y}[t]_{.i} \;\;
(i=1,...,q)$ it is necessary to construct a new vector of observations
$\mathbf{Y}^{*}_{.i}$ with new data and last parameter estimations, and
predictor matrix $\mathbf{X}^{*}$, and weight matrix $\bm{\Sigma}$ based on previous
variance parameters estimation as follows:
\begin{eqnarray}
\mathbf{Y}^{*}_{.i} &=& \left[
      \begin{array}{c}
        \mathbf{Y}[t]_{.i} \\
        \hat{\mathbf{W}}[t-1]_{.i} \\
      \end{array}
     \right]\\
\mathbf{X}^{*} &=& \left[
      \begin{array}{c}
        \mathbf{X}[t]\\
        \mathbf{I}_{p}\\
      \end{array}
     \right]\\
\bm{\Sigma} &=& \left[
      \begin{array}{cc}
        \mathbf{I}_{n}& \textbf{0} \\
        \textbf{0} & \mathbf{V}_{W}[t-1] \mathbf{s}^{2}_{i}[t-1] \\
      \end{array}
     \right]
\end{eqnarray}

New parameters estimation at time t could be written as:
\begin{eqnarray}
\hat{\mathbf{W}}[t]_{.i} &=& (\mathbf{X}^{*^\intercal } \bm{\Sigma}^{-1} \mathbf{X}^{*})^{-1}
\mathbf{X}^{*^\intercal } \bm{\Sigma}^{-1} \mathbf{Y}^{*}_{.i} \;\;i=1,...,q\\
\mathbf{V}_{W}[t] &=& (\mathbf{X}^{*^\intercal } \bm{\Sigma}^{-1} \mathbf{X}^{*})^{-1}\\
\mathbf{s}^{2}_{i}[t] &=& \frac{n_{0} \mathbf{s}_{0i}^{2}+n_1 \mathbf{s}_{1i}^{2}}{n_0+n_1}
\end{eqnarray}
where
\begin{equation}
s_{1 i}^{2}=\displaystyle \frac{(\mathbf{Y}[t]_{.i}-\mathbf{X}[t] \hat{\mathbf{W}}[t]_{.i})^\intercal (\mathbf{Y}[t]_{.i}-\mathbf{X}[t] \hat{\mathbf{W}}[t]_{.i})}{n_1-p}
\end{equation}
moreover, $s_0^2$ is the variance estimation at time $t-1$ and $n_0$ is its degrees of
freedom, and $n_1$ is the degrees of freedom in the new data variance
estimation. Computational cost is higher than first step, with more matricial
products and more inverse matrices calculus. The Bayesian standard parameters
estimation is a simple process but with high resources requirements.

\setcounter{equation}{0}
\renewcommand\theequation{B\arabic{equation}}
\section{On-Line Back-Propagation Derivation for ANN Models}\label{sec:bp}

This section mathematically formalizes the BP algorithm, which is widely known in
ANN related literature.~These equations are described for completeness and to help
the understanding of memory requirements stated at Section~\ref{sec:onlinebp}.

For any ANN model, with zero or more hidden layers, the procedure of computing
its output $\hat{\mathbf{y}}$ given its inputs $\mathbf{x}$ is known as
\emph{forward} step. The following equations show the computation needed during
forward step:
\begin{eqnarray}
  \mathbf{h}_{0} &=& \mathbf{x} \label{eq:forward1} \\
  \mathbf{h}_{j} &=& s( \mathbf{W}_{j} \cdot \mathbf{h}_{j-1} + \mathbf{b}_{j} ) \, , \text{ for } 1 \leq j < N\\
  \hat{\mathbf{y}} &=& \mathbf{W}_{N} \cdot \mathbf{h}_{N-1} + \mathbf{b}_{N}
  \label{eq:forward2}
\end{eqnarray}
being $s(z) = \frac{1}{1 + \exp(-z)}$ the logistic function and $N$
the number of layers in the network, that is, the number of hidden layers plus
one because of the output layer.~With forward step, all hidden $\mathbf{h}_j$
and output layer $\hat{\mathbf{y}}$ activations are computed. Following the
forward step, it is possible to compute the loss $L(\hat{\mathbf{y}}, \mathbf{y})$ of
the ANN output respect to the given desired output $\mathbf{y}$ by using the
mean square error.~The derivation of this loss function respect to every output
and hidden layer is computed by the \emph{backprop} step by means of the next
equations:
\begin{eqnarray}
  L(\hat{\mathbf{y}}, \mathbf{y}) &=& \frac{1}{2} \| \hat{\mathbf{y}} - \mathbf{y} \|_2^2 + \frac{\upepsilon}{2}\sum_{j=1}^{N} \sum_{\upomega \in \mathbf{W}_{j}} \upomega^2
  \label{eq:backprop1}\\
  \bm{\updelta}_{N} &=& \frac{\partial L(\mathbf{x},\mathbf{y})}{\partial \hat{\mathbf{y}}} = \hat{\mathbf{y}} - \mathbf{y}\\
  \bm{\updelta}_{j} &=& \frac{\partial L(\mathbf{x},\mathbf{y})}{\partial \mathbf{h}_j} = \mathbf{h}^\prime_{j} \circ \left( \mathbf{W^\intercal}_{j+1} \cdot \bm{\updelta}_{j+1} \right)\, , \text{ for } 1 \leq j < N
  \label{eq:backprop2}
\end{eqnarray}
being $\mathbf{A} \circ \mathbf{B}$ the component-wise product between
two vectors and $\mathbf{h}^\prime_{j}$ the derivative of the logistic function,
that corresponds to $\mathbf{h}^\prime_{j} \circ \left( 1 -
\mathbf{h}^\prime_{j} \right)$. After the forward and backprop steps, all
activations and hidden and output layer error gradients $\bm{\updelta}_j$ are
available. Thus, the gradient of the loss function respect to the weight matrices
and bias vectors can be computed. Following this gradient, all weights
and biases are updated. Such phase is denoted by \emph{update} step, and
its equations are:
\begin{eqnarray}
  \mathbf{W}^{(e+1)}_{j}
  &=& \mathbf{W}^{(e)}_{j} - \upeta \frac{\partial L(\mathbf{x},\mathbf{y})}{\partial \mathbf{W}^{(e)}_{j}}\label{eq:update1} =
  \mathbf{W}^{(e)}_{j} - \upeta \cdot \left( \bm{\updelta}_{j} \otimes \mathbf{h}_{j-1} + \upepsilon \mathbf{W}^{(e)}_{j} \right)  \, , \text{ for } 1 \leq j \leq N\\
  \mathbf{b}^{(e+1)}_{j}
  &=& \mathbf{b}^{(e)}_{j} - \upeta \frac{\partial L(\mathbf{x},\mathbf{y})}{\partial \mathbf{b}^{(e)}_{j}} =
  \mathbf{b}^{(e)}_{j} - \upeta \cdot \bm{\updelta}_{j}  \, , \text{ for } 1 \leq j \leq N
  \label{eq:update2}
\end{eqnarray}

\noindent being $\mathbf{A} \otimes \mathbf{B}$ the outer product between two
vectors.

\conflictofinterests{Conflicts of Interest}

 The authors declare no conflict of interest.

%


\end{document}